\documentclass{JHEP3}
 \pdfoutput=1

\usepackage{graphicx}
\usepackage{amsmath,amsfonts}

\numberwithin{equation}{section}

\DeclareMathOperator{\Tr}{Tr}
\DeclareMathOperator{\erf}{erf}
\DeclareMathOperator{\SU}{SU}
\renewcommand{\phi}{\varphi}
\renewcommand{\epsilon}{\varepsilon}
\DeclareMathOperator{\C}{\cal C}
\renewcommand{\O}{\rm O}

\newcommand{\R}{\mathbb R}

\title{\boldmath Rectangular Wilson Loops at Large N}

\author{R.~Lohmayer${}^{(a)}$ and H.~Neuberger${}^{(b)}$ \\
Rutgers University, Department of Physics and Astronomy, Piscataway, NJ 08854,
USA
${}^{(a)}$\email{lohmayer@physics.rutgers.edu}\\
${}^{(b)}$\email{neuberg@physics.rutgers.edu}\\
}

\abstract{
This work is about pure Yang-Mills theory in four Euclidean dimensions with gauge group $\SU(N)$.
We study rectangular smeared Wilson loops on the lattice at large $N$ and  
relatively close to the large-$N$ transition point in their eigenvalue density. 
We show that the string tension can be extracted from these loops but 
their dependence on shape differs from the asymptotic prediction of effective string theory. 
}


\begin{document}

\section{Introduction}

This article is about Wilson loop operators, $W(\C)$, in four-dimensional
Euclidean $\SU(N)$ pure gauge theory. $\C$ is a closed, non-self-intersecting 
continuous curve in
$\R^4$, differentiable except at a finite set of points where there 
are finite discontinuities in the unit tangent vector to the curve. $W(\C)$ is
well defined only after a renormalization that eliminates perimeter and corner
divergences. 

The inclusion of kinks is a complication one would like to avoid in initial studies.
It becomes essential if one wants to make the Makeenko-Migdal equations well defined.
It also is needed if one wishes to employ lattice field theory tools to validate non-perturbative 
assumptions about $W(\C)$ for contractible $\C$.

\subsection{Smearing}

A convenient way to renormalize $W(\C)$ is to use continuum smearing, 
henceforth referred to as ``smearing''. Smearing is 
a well defined procedure in Euclidean continuum field theory, abstracting some
more ad-hoc procedures in common use in lattice field theory. 
It was introduced in \cite{nn-smear} and a brief review of its history
can be found in \cite{ln-lat2011-smear}.  Smearing introduces an extra parameter,
$s$, of dimension length squared. $\sqrt{s}$ is an observer's resolution 
of localized objects constructed out of fields. Keeping this resolution
nonzero eliminates all divergences associated with operator compositeness. 
Smearing enjoys several useful properties:
\begin{itemize}
 \item All standard general regularization methods (purely perturbative or lattice) 
 are compatible with smearing. Perturbatively, 
 the counter terms to the classical Lagrangian required by ordinary renormalization 
 make all observables constructed out of only smeared fields finite -- with no
 restriction on their space-time arguments. Beyond perturbation theory, 
 choosing some reasonable definition of a scale eliminates ultraviolet divergences from 
 all smeared observables. 
 \item $\O(4)$ spacetime invariance remains preserved if the regularization 
 preserves it.
 \item Gauge invariance remains preserved if the regularization preserves it. 
 \item For any $\C$, smearing provides a proper definition of the joint
 eigenvalue distribution of parallel transport round $\C$ and these eigenvalues
 reside on the unit circle. The concept of a marginal probability distribution
 for the parallel transport unitary matrix round $\C$ in the continuum limit
 makes sense only after smearing. The classical view of parallel
 transport as an element in the compact $\SU(N)$ group is preserved 
 at the quantum renormalized level thanks to smearing, but won't hold with 
 more standard methods of continuum regularization. 
 In particular if we allow $\C$ to have an exactly backtracking segment,
 subsequent passages will cancel exactly out at the quantum level. 
 Smeared Wilson loops obey Polyakov's \cite{zig-zag} zigzag symmetry.
 \item At the formal level of the Makeenko-Migdal loop equations, 
 the definition of smearing can be extended to loop space \cite{ln-prl}. 
 In this formal sense, smearing is defined without any reference to the 
 gauge fields and their Lagrangian. The smearing parameter plays the role
 of the evolution variable in a generalization of diffusion to loop space. 
 \item Small smeared Wilson 
 loops admit a local expansion \cite{ln-lat2011-smear} 
 with well defined ``condensates'' at each order.
 These condensates can be rendered dimensionless by multiplication by a power
 of $s$. These dimensionless numbers provide non-perturbative definitions of 
 ``running'' coupling constants at scale $s$. Unlike their phenomenological
 progenitors~\cite{svz},~\cite{wl-shifman}, 
 the condensates are well defined. They have a calculable 
 perturbative expansion. The terms in this expansion do not vanish. 
 Another way to define ``condensates'' 
 is from the large-momentum structure of a specific 
 observable \cite{ovlap-alphas-japanese}. 
 Such a definition does not ensure that precisely the same condensate  
 enters in other observables.  The smeared versions are defined in a manner independent
 of the observable. They are dependent on the smearing
 parameter. This parameter is denoted by $s$ in the continuum and by $S$ on the lattice. 
\end{itemize}

Admittedly, smearing is an artificial device. We do not know of an 
alternative construction of a full set of functionals $W(\C)$ 
with properties listed above.  

Preserving the full $\O(4)$ symmetry of space-time in the process of smearing has
the drawback that smeared correlation functions will not exhibit the unitarity  of
the underlying theory in a transparent manner. For those quantities that have a finite limit 
as $s\to 0$ transparency will be recovered. One can define a version of smearing
which only preserves an $\O(3)$ subgroup of $\O(4)$ and thus keep unitarity evident. 
The cost is the loss of explicit Euclidean $\O(4)$ invariance. 

It is important to keep in mind the distinction between the (continuous) smearing we 
are using here and more standard procedures. For example, in~\cite{necco_sommer} 
smeared rectangular Wilson loops were used to extract the interquark potential and ultimately force.
The smearing was restricted to three directions perpendicular to the ``time'' direction, which
was taken to infinity. The smearing was done by iteratively adding one-plaquette-windowed paths 
to the spatial portions of the loop with 
a weight of 1/2. 

Thus, this older version of smearing differs on two counts from the continuous smearing 
we are employing in this paper: It is only $\O(3)$ invariant and it is defined only on the 
lattice. 

The proper way to think about this smearing is as a procedure to improve the
numerical quality of the string energies defined in the asymptotic large-time regime. This 
smearing is a purely lattice technique and plays no role in the subsequent extraction of 
physically relevant parameters. No remnant of this smearing is left in the continuum limit.

The smearing we use in this work is intentionally 
designed to make the Wilson loop average itself
well defined in the continuum, rather than just the interquark force. To achieve this an
extra, tunable, dimensional parameter is introduced, intuitively describing the resolution at
which the Wilson loop is observed. 

\subsection{Weakly versus strongly coupled regimes}

In 4D $\SU(N)$ pure gauge theory, classical scale invariance is anomalous 
and gets broken at the quantum level. A 
scale separating a weakly coupled short distance regime  from a 
qualitatively different strongly coupled long distance regime is dynamically 
generated. Observables admit an asymptotic expansion computable in
perturbation theory at short distances. There are no systematic methods of
analytic computation at large distances. 
These two regimes coexist in the same theory and are smoothly connected. 
The crossover is relatively narrow. 

Numerical simulations have established that at large distance the theory confines. 
This is in agreement with experimental data for $N=3$.  A strong version
of the confinement postulate is: If we scale a loop $\C$ up, $\C \to \rho \C$, 
$W(\rho \C)\sim \exp(-\rho^2 \sigma A_{\rm min} )$ as $\rho\to\infty$ 
with $A_{\rm min}$ the minimal area 
of all continuous surfaces in $\R^4$ bounded by $\C$. 
The string tension $\sigma$
has dimension of mass squared and is universal: It does not 
depend on the shape of $\C$.
The numerical and the empirical evidence are both restricted to simpler
loops. Typically, they fit in a plane inside Euclidean $\R^4$,  are 
non-self-intersecting and not too unusually shaped. 

The string tension can be extracted from unsmeared  
Wilson loops $W(\C,s=0)$ by numerical means. The externally determined 
resolution scale $\sqrt{s}$ can be thought of as representing  
an effective thickness of $\C$. For very large loops the fixed finite thickness 
$\sqrt{s}$ should not enter. One expects 
$W(\rho \C, s)\sim  \exp(-\sigma \rho^2 A_{\rm min}(\C) )$ 
with $\sigma$ independent of $s$. 

There does not exist so far a mathematically rigorous proof of confinement in the
continuum limit. Even if we postulate confinement, 
there is no credible analytical computation of 
$\sigma$ in terms of a perturbatively defined scale $\Lambda$. 
We think methods of effective field/string theory could achieve this.
In this paper we use the lattice to study in detail 
the crossover from weak to strong coupling. 
We hope to learn how to quantitatively 
connect the confinement regime to the weakly coupled 
one and eventually estimate $\sigma/\Lambda^2$ 
without using the lattice in any quantitative way.

The spirit is the same as in a 
semiclassical approximation based on instantons \cite{cdg-mit}. 
That approach 
was not successful but framed the problem well. It tried to connect the 
perturbative regime to a regime 
in  which an MIT bag description held by a crossover 
described using an instanton gas. A 
more phenomenological approach 
was based on an extension of perturbative 
OPE to include nonperturbative contributions parametrized 
by ``condensates'' \cite{svz}. This approach was quite successful but is
imprecisely defined. It remains unclear how a theorist would extract 
an exact value of a universally meaningful ``condensate'' 
even if she/he somehow managed to solve
QCD exactly in the presence of an acceptable UV cutoff.  

Perhaps it is not by accident that our concrete approach 
employs an artificial smearing scale.
Smearing provides well defined candidates for SVZ condensates, as already mentioned. 
These condensates also are not directly
physical since smearing is somewhat ad-hoc. They are universally meaningful though.
They would be quantitatively useful only if one chose a reasonable 
level of smearing. A good choice would provide an economical parametrization of the short 
distance -- long distance crossover. 

This paper is part of a general strategy. 
We want to gain analytical control of the crossover for simple Wilson loops
by exploiting newly established large-$N$ phenomena by computer simulation. 
Then we want to compute $\sigma$ in units of a perturbative scale 
$\Lambda^2$. Confinement is assumed and one accepts 
an effective description of the confinement regime 
based on the scale $\sigma$. This might
become practical long before a mathematical proof of confinement 
is found and without a detailed understanding of what causes it. 

\subsection[Large $N$]{\boldmath Large $N$}

The essence of the weak -- strong coupling problem remains present 
in the limit $N\to\infty$. 
The limit is taken as $N g_N^2 (s) \equiv\lambda(s)$ is kept fixed 
\cite{thooft-large-n}. 
$g_N^2 (s)$ is a running coupling constant in some standard definition.

It has been recently established that the weak-strong 
crossover range collapses into a well defined
point at infinite $N$ \cite{ln-prl}. 
The transition point depends on the shape of $\C$. 
As the loop $\C$ is dilated at fixed non-zero
smearing and shape a non-analytic change in the single-eigenvalue distribution takes place
at a sharply defined scale. For a small loop the distribution is
insensitive to the compact nature of $\SU(N)$. For a larger loop the
full group is explored by the parallel transporter round it. 
For finite $N$ there is no non-analyticity and the full group is felt 
by parallel transport around all loops. 

Group
compactness is a key ingredient for confinement. Perturbation
theory is insensitive to it because it starts from an infinite-range Gaussian
integral over YM fields. The transition at which
the eigenvalues of the parallel transport matrix ``discover'' the point
in the group that is farthest from identity is a natural scale 
for matching perturbation theory to a long distance description. 
Traces of smeared Wilson loops
in the fundamental representation remain smooth through the transition 
even at infinite $N$, although the single eigenvalue distribution is
not analytic there. 
Therefore, traces of smeared Wilson loops should match well. 
The small loop regime is in principle
calculable by field theory. More recently we have learned that one can 
also make predictions by analytical means about the large loop regime. 
The framework for doing that is effective string theory.

\subsection{Effective string theory}

Effective string theory \cite{aharony}, \cite{eff-string-th} bears conceptual similarity to the well known 
chiral effective field theory describing the interactions between soft
pions in an $\SU(N)$ gauge theory with a moderate number of massless
quarks. One assumes that spontaneous chiral symmetry
breaking occurs via a bilinear condensate 
and that the finite non-zero pion decay constant $f_\pi$ is
the scale typical for effects caused by this breaking. Then, 
symmetry considerations produce a large set of predictions. 
One has a proof \cite{thooft-chiral} for chiral symmetry breaking at a
physicist's level of rigor but the proof provides no indication for how to 
calculate $f_\pi$ in terms of $\Lambda$ by analytical means. 
Consider the correlation
function of two flavor currents in QCD. We have a perturbative description
at short distances and a chiral effective field theory description at long
distances \cite{svz}, \cite{derafael}. Joining them at a crossover scale would provide 
some estimates for the ratio $f_\pi^2/\Lambda^2$. Resonance 
contributions come in in the crossover regime and the match is complicated. 
The large-$N$ limit simplifies matters somewhat because the resonances 
become isolated stable particles coming in as poles.

The Wilson loop analogy is substantially less developed and we think that time
has come to look into the problem 
of matching short to long distances for Wilson loops in some detail. 
The existence of the sharply demarcated matching point on the one
hand and the smoothness of the observable through this matching
point on the other are encouraging. 

We want to determine by lattice gauge theory methods
how an effective string description on the strong coupling side of the
matching point and close to it works in detail. 
The effective string theory makes predictions for 
a functional of curves $\C$ and
the central assumption is that these predictions describe Wilson loops. 
The string tension $\sigma$ is used to set the scale in the theory from the outside.
The effective string theory in itself does not generate any scale. 
An important issue is how this matching depends on scale-invariant features 
of $\C$. 

The predictions are obtained starting from a limit where the 
minimal spanning area of $\C$ is very large. 
One can ignore any length scale that stays fixed as $\C$ is dilated. 
Very large loops are described by the Nambu-Goto
action for 2 massless bosonic fields. It is impossible to define this limiting 
theory exactly. This is not needed as 
an asymptotic expansion in inverse loop-size 
produces well defined terms 
without a full definition of the theory. 
The predictions made by  
effective string theory are obtained from the expansion of the action
around its quadratic approximation. Eventually
one reaches an order at which non-Nambu-Goto terms are needed. 
Unlike in the chiral Lagrangian case, this order appears to be relatively high. 
The main ingredient organizing the expansion is the postulated local nature of the 
world-sheet theory. Effective string theory for contractible Wilson loops and 
effective field theory for massless pions differ in scope. The effective chiral 
Lagrangian applies to functions of a finite number of scales, while effective 
string theory applies to a functional of a continuum of scales. 

String theory would require the inclusion of handles in the
calculation of corrections. It is believed that
handles can be neglected at infinite $N$.  Thus, in the 't Hooft limit, 
one ends up using just purely field-theoretical methods
of two-dimensional field theory when one imposes 
on the effective string description
symmetry restrictions coming from the original theory. 
 
In this numerical work we do not have data of quality needed to identify 
terms predicted by the effective string approach beyond the determinant 
of Gaussian surface fluctuations. The leading 
term states that as the loop is dilated to infinite size, at fixed shape, 
$\log W(\C, s)\sim -\sigma A_{\rm min} (\C)$. 
The two next subleading terms in the
asymptotic expansion around very large loops come from the determinant
of small fluctuations of the surface bounded by $\C$ about its absolutely minimal
area configuration. This configuration is assumed unique and well separated
from other minima. The first subleading term
is proportional to $\log(A_{\rm min})$ and the second is invariant under 
scalings of $\C$.  Except for simple contours, it won't be possible to
write down explicit formulas for the second subleading term, but, 
for any specific $\C$, the value of the term can be obtained numerically
with relative ease. 

The effective string theory 
cannot make predictions for two terms that are present in $\log W(\C, s)$ 
and come in between its leading and its subleading predictions. These
unpredictable terms consist of a perimeter term and a corner term. Both
are smearing dependent.  A potential problem
then arises of an ``interference'' 
between further 
smearing-dependent subleading terms in $W(\C, s)$ and smearing independent terms
coming from the effective string theory. The consequence  of 
this ``interference'' is that even
given an exact formula for $W(\C,s)$, coming from $\SU(N)$ field theory, we would
not be able to check whether the effective string theory works or not 
because we would not be able to separate out the effective-string prediction. 
The higher-order effective-string predictions could then just ``melt away'' 
into the exact expression. 

The field theory produces a $\log(W(\C,s))$ which diverges as $s\to 0$. 
The divergences appear in 
an asymptotic expansion at fixed $\C$ in $s$ as $s\to 0$. Subtracting
the $\C$-dependent terms that go as $\frac{1}{\sqrt{s}}$ and 
$\log^\kappa {s\Lambda^2}$ would yield a finite expression. 
$\kappa=1$ at tree level in perturbation theory and subsequent terms 
in the leading log approximation can be resummed using the Callan-Symazik equation. 
This produces a term going like $\log(\log s\Lambda^2)$. 

Let us consider the case of a very small smearing parameter first: $s\sigma \ll1$. 
It makes sense now to just subtract terms that diverge as $s\to 0$ and then set $s=0$.
This would produce as ``pure'' a Wilson loop as any other regularization, 
which does not involve smearing, would. 
The power divergence is a perimeter term, well separated 
from other terms. It is clear how to subtract it without introducing finite terms
that depend on the curve. There are no logarithmic divergences proportional
to the perimeter. The subtraction of the logarithmic
divergence depends on the opening angles of the corners. 
It is unclear how to disentangle the remaining finite parts from the subtraction of the corner 
divergences from unrelated terms dependent on 
shape features of $\C$. By definition we are considering an asymptotic expansion 
in a parameter describing by how much the Wilson loop has been dilated 
relative to a standard size. The coefficients in the asymptotic 
expansion in the inverse of the dilation parameter are nontrivial functions of the shape
of the loop, which is kept constant throughout. Effective string theory can be used to
provide expressions for these coefficients. These expressions consist of functions of loop
shape, but not overall scale. This particular asymptotic expansion differs from other
variants, which also are produced by effective string theory. For example, one might
extract the interquark force from rectangular loops and consider the expansion of
this force in the distance between the quarks measured in units 
of the string tension, $\sqrt{\sigma}R$. Now the coefficients are just pure
numbers.

In our context we are left with an open question 
as to what effective string theory predictions for the dependence on 
shape parameters of $\C$ should be compared to. 
One of our objectives in this paper is to get some 
guidance on this question from numerical simulation.

\section{Rough outline of paper}

We have obtained Monte Carlo estimates for smeared rectangular
Wilson loops on a hypercubic lattice at various smearing levels, 
$N$'s, couplings, volumes and combinations of rectangle sides. The
estimates were obtained using a data base of 160 uncorrelated 
equilibrated gauge fields  we have distributed on a forty node PC cluster. 
Each cluster node has four cores and a total of 24GB of memory to be able to 
smear and make measurements on four distinct gauge fields simultaneously. 

Wilson loops in all distinct orientations and
locations were averaged over for each gauge configuration separately. 
For each set of parameters defining
the gauge field action and the loop we obtain 160 numbers. The set of these 
numbers is used for the statistical estimation of various physical parameters.
Statistical errors are always determined by jackknife with the 
elimination of one single gauge configuration from the set of 160 at a time. 

We start by extracting the string tension from square loops. The infinite-volume
and large-$N$ limits on the lattice are dealt with first. Then the string tension is
extrapolated to its continuum limit. The results and extrapolations are validated 
against a set of $L \times  L\!+\!1$ and $L\times 2L$ loops.
This analysis is done assuming that the term logarithmic in the area
is precisely the one predicted by effective string theory but making
no assumptions about the shape-dependent terms. This is achieved 
by ensuring  that shape-dependent terms play no role at this stage. 
Under the same assumption about the logarithmic area dependence 
a loop-shape dependent number is extracted
from the data and compared to the string prediction. 
The data is revisited and using global fits the coefficient 
of the logarithm of the area is determined. 
We obtain numbers consistent with the effective string theory value.
This validates the assumption we made earlier when the coefficient
was held fixed. The dependence on smearing is addressed throughout. 

\section{Parameter choices}

We use the standard Wilson single-plaquette action. The standard  coupling $\beta$
is incorporated into $b=\frac{\beta}{2N^2}$. The $N\to\infty$ limit is taken at fixed $b$. 
All our gauge configurations are on symmetric
hypercubes of side $V^{\frac 1 4}$. $V$ is the total number of sites and
the boundary conditions are periodic. We wish to use effective string theory
to understand the data and the structure of the latter is restricted in the
continuum by target space $\O(4)$ invariance. In order to maintain as much
of the latter as possible at the regularized level we work only with symmetric volumes.

For large $N$ there is a bulk transition close to $b=0.360$. 
For substantially smaller values of $b$ the system is in a phase disconnected from
continuum Yang-Mills theory. 
We also need to maintain $b\leq 0.369$ to be sure that spontaneous 
$Z^4(N)$ breaking at $N=\infty$~\cite{plb-red} is avoided on all our volumes, 
including our smallest, $12^4$.  We mainly use the range of couplings 
$0.359 \le b \le 0.369$. These couplings can produce relatively small and fine lattices. A judicious 
exploitation of continuum large-$N$ reduction allows us to always carry out the required extrapolation 
to infinite volume. Continuum large-$N$ reduction is sometimes also referred 
to as partial reduction. It is a conservative version of reduction introduced in~\cite{prl-red}. 
Henceforth we shall use the term ``reduction'' instead of ``continuum reduction''. 

We stored statistically independent 
gauge fields at intervals $\Delta b=0.001$. Satisfactory statistical 
independence for our observables is obtained for gauge fields at neighboring
$b$'s being separated by 500 complete $\SU(2)$ updates combined with
500 complete over-relaxation passes.\footnote{A complete $\SU(2)$ update 
consists of sequential updates of $\frac{1}{2} N(N-1)$ $\SU(2)$ subgroups. 
Similarly, a complete overrelaxation update consists of a ``reflection''
of the entire $\SU(N)$ link matrix.}
 The autocorrelation time is equal to about 
one quarter of this separation. 
The set of $N$ values we use consists of $N=7$, $11$, $13$, $19$, $29$. 
The computer time for generating a gauge field configuration goes as $N^3 V$ and this
is the primary limitation on the $(N,V)$ combinations we use. 

Each measurement proceeds after the gauge fields have been smeared. 
The smearing parameter $S$ has mainly been taken in the range $0.2 \leq S \leq 0.4$.
In some cases we have data up to $S=0.52$. The separation between sequential smearing 
levels is $\Delta S=0.04$. 

The Wilson loops $W_N$ on the lattice are defined by 
\begin{align}
W_N (L_1,L_2,b,S,V)=\frac 1N \langle \Tr  \prod_{{\it l}\in \C} U_{{\it l}} \rangle\,.
\end{align}
The product is over the links ${\it l}$ in the order they appear when
one goes once round $\C$, a rectangle of sides $L_{1,2}$.   All our
fits will be applied to 
\begin{align}
w_N(L_1,L_2,b,S,V)=-\log W_N(L_1,L_2,b,S,V) \,.
\end{align}
When the loops are square the two variables $L_1,L_2$ are replaced
by one $L$ with the understanding that $L_{1,2}=L$.

\section{Square loops}
\subsection[$N\to\infty$ and $V\to\infty$ limits]{\boldmath $N\to\infty$ and $V\to\infty$ limits}
\label{Sec:NVlimits}

We first want to determine the limit
\begin{align}\label{eq:limNlimVofW}
\lim_{N\to\infty}\left (\lim_{V\to\infty} w_N(L,b,S,V)\right )\,.
\end{align}
Numerically this is nontrivial since we need a good level of
accuracy on the limit. Fits extracting physical parameters 
are applied to estimates for this limit. 

Once $V$ is larger than some moderate $V_c(b)$, large-$N$ reduction provides 
in principle a shortcut allowing one to drop the 
limit $V\to \infty$ above. At which $N$ rough convergence is
attained will depend on $V$. This requires numerical tests and fits.  
In addition to $N$, $V$, and $b$ the magnitude of finite-volume corrections also depends on 
the observable: the larger it is the bigger the finite-volume effects that
need to be overcome are. 

We use two different methods to 
compute the limit \eqref{eq:limNlimVofW}:

\begin{itemize}
\item Method 1) 
\newline
At fixed $N$ we compute $w_N$ on volumes that are sufficiently large for
finite-volume effects to be negligible, then we determine
$w_{\infty}(V=\infty)$ by 
fitting $w_N(V=\infty)$ to 
\begin{align}\label{eq:infNinfVfit}
w_N(V=\infty)=w_\infty(V=\infty)+\frac{a_1(V=\infty)}{N^2}+\frac{a_2(V=\infty)}{N^4}\,.
\end{align}
Here, the other arguments of $w$ are omitted for simplicity.
All coefficients depend on the observable. 
\newline
We have evidence (strong for $N=7$ and $N=11$, not that strong for
  $N=19$ and rather weak for $N=29$) 
  that volumes $V=24^4$, $18^4$, $14^4$, $12^4$ 
  are sufficiently large for $N=7$, $11$, $19$, $29$, respectively.  
This statement applies to the specific set of couplings and loop sizes we use. 
The evidence comes from comparing to results on other volumes $V$ and/or 
other values of $N$ or even from trying to extrapolate from lower values of $b$. 
The accuracy of our data does not allow to quantify the finite-volume effects and
we have to settle for something more qualitative. We managed to 
convince ourselves that 
the finite-volume systematical deviations are smaller than our statistical errors. 

\item Method 2) 
\newline
The second method makes use of large-$N$ reduction. At fixed $V$, we first
take the limit $N\to\infty$ of $w_N(V)$ by fitting
\begin{align}\label{eq:infNfixedVfit}
w_N(V)=w_\infty(V)+\frac{a_1(V)}{N^2}+\frac{a_2(V)}{N^4} \,.
\end{align}
So long as the center symmetry stays unbroken, 
there is no volume dependence in the infinite-$N$ theory, i.e.,
$w_\infty(V)=w_\infty(V=\infty)$.
\newline 
We determine $w_\infty(V=12^4)$ 
from $N=11$, $13$, $19$, $29$ [method 2a)] and
$w_\infty(V=14^4)$ from $N=7$, $11$, $13$, $19$ [method 2b)].     

There is little theoretical doubt that reduction indeed holds as a statement 
about $N=\infty$ for values of $(b,V)$ in the allowed region. The limit
$N\to\infty$ is unlikely to be uniform in $(b,V)$ or in
the size of the loop $\C$ and in the level of smearing $S$.  If we see good fits 
to a sum of terms decreasing as $\frac{1}{N^2}$ we know that we have 
taken into account subleading effects that do have a dependence on $V$. 
Only then can we trust that the leading term is indeed $V$-independent. 
\end{itemize}

\FIGURE[htb]{
\includegraphics[width=0.8\textwidth]{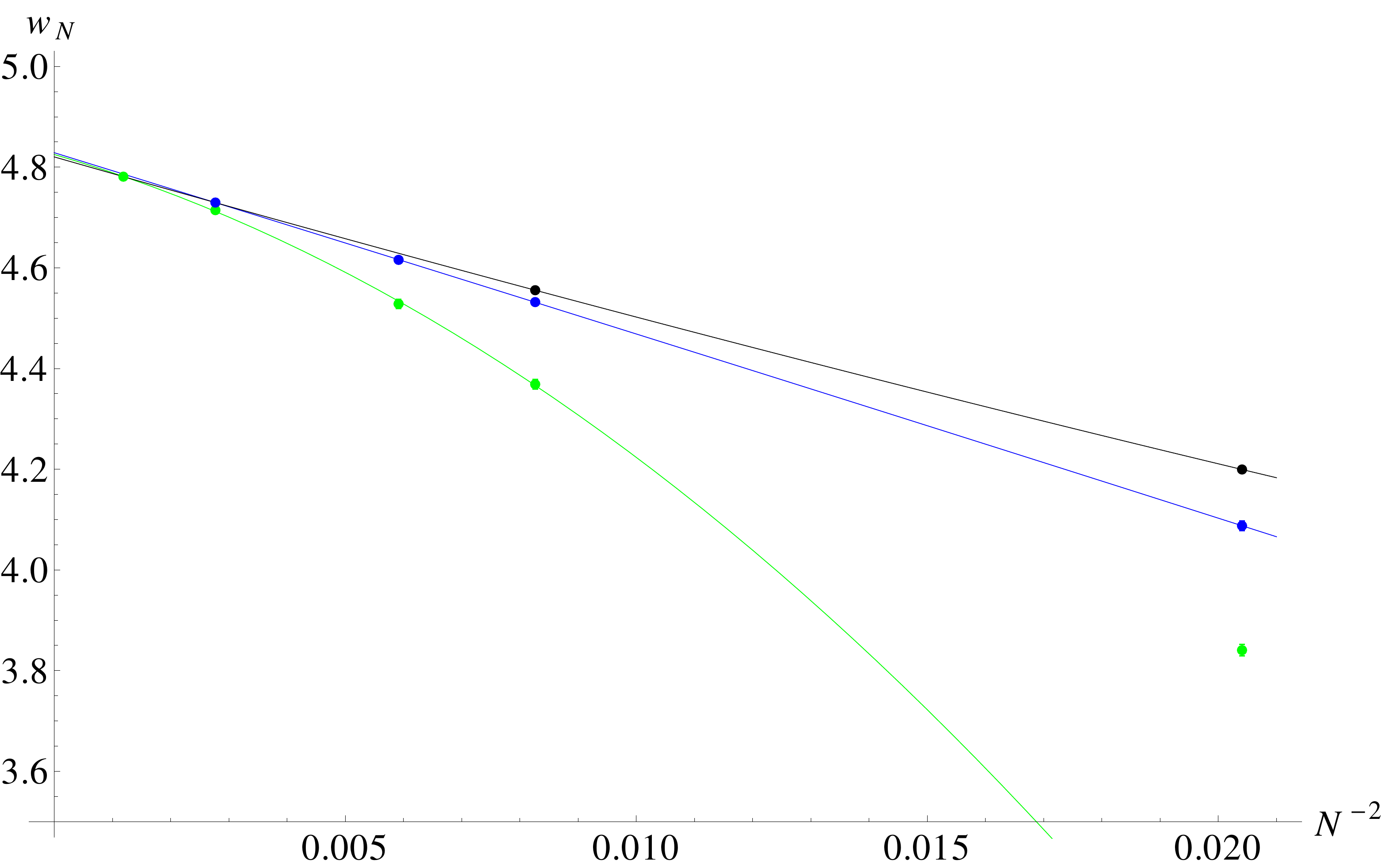}
\caption{Plots of $w_N(L=9,b=0.368,S=0.4,V)$ as a function of $1/N^2$: $V=12^4$
  in green, $V=14^4$ in blue and $V=24^4$ (at $N=7$), $V=18^4$ (at $N=11$) in
black. We display fits for each method: 1) by a black solid line, 2a) by a 
green solid line and 2b) by a blue solid line. Error bars are not visible in the plot.} 
\label{fig:infNExample} 
}

Figure~\ref{fig:infNExample} shows an example for the three different
extrapolation methods at $b=0.368$, $S=0.4$, $L=9$. 
Comparing the results we obtain for
$\lim_{N,V\to\infty} w_N(L,b,S,V)$ in the three different methods 1), 2a), 2b)
we obtain credible numbers for $w_\infty(\infty)$. Some comments are in order:
\begin{itemize}
\item We obtain reasonable values of $\chi^2/N_{\text{dof}}$ for the fits \eqref{eq:infNinfVfit} and
  \eqref{eq:infNfixedVfit} for 1), 2a), 2b).  2a) is an exception where we have
    $\chi^2/N_{\text{dof}}$ up to 6 for $b\leq 0.361$. We find 
    good agreement between the results for
  $\lim_{N,V\to\infty} w_N(V)$. This agreement is compatible with their statistical accuracy of
  about 0.1\%.   The worse $\chi^2/N_{\text{dof}}$ at $b\leq 0.361$ likely reflect the 
  impact of the $N=\infty$ bulk transition in the respective volumes.  We keep this in mind
  in subsequent analyses. 
\item We cannot determine the coefficients $a_2$ very accurately. 
Truncating the expansions \eqref{eq:infNinfVfit} and
  \eqref{eq:infNfixedVfit} at $\mathcal O(N^{-2})$ would result 
  in very large $\chi^2/N_{\text{dof}}$, so $a_2$ cannot be set to zero. 
\item Including the $N=29$ result in the fit
  \eqref{eq:infNfixedVfit} for $V=12^4$ is crucial for 2a) to agree with 2b) and
  1) for large loops and large $b$. Including $N=7$ in the $V=12^4$ fit 
  would require an additional $1/N^6$
  correction in \eqref{eq:infNfixedVfit}. See~Fig.~\ref{fig:infNExample} for
  an example. For volumes close to minimal, there is no useful information 
  to be gained about the $N,V=\infty$ limit 
  from numbers obtained at low values of $N$. 
\end{itemize}
When the lattice size $V^{\frac 14}$ is getting close to the
critical lattice size $L_c(b)$ at which the center symmetry brakes, 
we need to go to higher $N$'s if we want to
compute $\lim_{N,V\to\infty} w_N(L,b,S,V)$ using method 2).
The required computation time scales as $N^3V$. 2a) is about
1.75 times more expensive than 2b) and 1) is about 2.5 times more expensive
than 2b). It is hardly possible to conclude from 2b) or 2a) alone that the estimates
for $w_\infty(V)$ are reliable. We became confident that we have correctly determined
$\lim_{N,V\to\infty} w_N(V)$ only after having obtained agreeing results
from 1), 2a) and 2b).

\subsection[Lattice string tension at infinite $N$]{\boldmath Lattice string tension at infinite $N$}
\label{sec:infNSigma}

At fixed smearing level $S$ and coupling $b$ we  use the shorthand notation: 
\begin{align}
w_\infty(L)\equiv\lim_{N,V\to\infty} w_N(L,b,S,V)\,.
\end{align} 
For square $L\times L$ loops, we expect
\begin{align}\label{eq:wLL}
w_\infty(L)+\frac 14 \log L^2 = c_1 + c_2 L + \sigma L^2 + \mathcal
O\left(\frac 1{\sigma L^2}\right)\,.
\end{align}
The log term comes from the determinant of small fluctuations around 
the minimal area configuration in the effective string description. 
We shall return to it later. For now, its presence is just assumed 
because including it gives good fits while excluding it gives bad fits.

Neglecting corrections of order $\frac 1{\sigma L^3}$, we fit
\begin{align}\label{eq:SigmaFromSquares}
\frac 12 \left(w_\infty(L+1)-w_\infty(L)+\frac 12 
\log\left(1+\frac 1 L\right)\right)=\sigma \left(L+\frac 12\right)+\frac{c_2}{2}+ \mathcal
O\left(\frac 1{\sigma L^3}\right)
\end{align}
to a straight line as a function of $L+\frac 12$ to determine $\sigma$ and
$c_2$. See Fig.~\ref{fig:SigmaExample} for some examples. Subsequently 
we use this determination to subtract the area and the
perimeter term from the data and fit $w_\infty(L)+\frac 14 \log L^2-\sigma
L^2 - c_2 L$ to a constant, $c_1$. We carry out the fits in this order 
because the numerical contribution to $w_N$ 
of $\sigma L^2+c_2 L$ substantially 
exceed that of $c_1$. This will be shown later. 
The numerical contribution 
of the perimeter term is large because it  
would diverge like $s^{-\frac 12}$ as $s\to 0$. Allowing too large a perimeter term 
has the negative effect of reducing $W_N$ and consequentially decreasing the 
signal to noise ratio. Reduced smearing increases the statistical error. 

Most  $5\times 5$ loops fall
into the neighborhood of the large-$N$ phase transition in the
eigenvalue spectrum of the Wilson loop matrix for the $b$ and $S$ values we work with. 
Physically smaller loops will have a single-eigenvalue distribution which
has a gap around -1. 
We expect confinement to have something to do with the compactness of 
$\SU(N)$ after it is dynamically detected. We know this for a fact in 
the exactly soluble case of two-dimensional YM~\cite{loop-instantons}.
In 2D one can analytically separate out all contributions 
depending on eigenvalues exploring the entire unit circle. 
The remainder of the expression for $W_N$ no longer 
exhibits an area law. Demonstrating the separation requires one to
first introduce an infrared cutoff because perturbative and non-perturbative contributions 
mix at infinite volume. The absence of confinement has been shown on a sphere, a cylinder 
and at infinite volume by using the 
Wu-Mandelstam-Leibbrandt IR regularization~\cite{inst-two-d}.

Therefore, for effective string theory fits we
use only loops with $L\geq 6$ to determine the parameters $\sigma$, $c_2$, and
$c_1$. Wilson loop matrices for such loops 
have a gapless eigenvalue spectrum
even in the infinite-$N$ limit.  
The results presented below are obtained from square loops in the range
$6 \leq L \leq 9$. 
Best fit parameters are obtained by using weighted least-square fits. 

Figure~\ref{fig:Sigma-b} shows results obtained for $\sigma$ using the
different methods for computing $w_\infty$ for $0.359\leq b \leq 0.369$ at smearing level
$S=0.4$. The three results agree with each other within statistical errors. These 
errors are smallest for method 1). 

Decreasing the smearing level to $S=0.28$ results in increasing statistical
errors for $\sigma$. Within these errors, $\sigma$ does not depend on $S$, 
cf.~Tables~\ref{tab:sigmas} and \ref{tab:sigmasS14}.  This is as we expected and indicates that
one can use smeared loops as a device to compute the string tension. In this sense,
the numerical benefit of the type of smearing used before 
continuous smearing was introduced,
~\cite{necco_sommer} is maintained. In the older version there is no 
need to test whether 
the extracted string tension depends on smearing because the 
infinite time limit projects on lowest energy states that by definition 
are independent of the details of the projector and this
version of smearing only affects the latter. 
When one extracts the string tension from finite rectangular loops, the independence on 
the continuous smearing parameter should be checked.

\FIGURE[htb]{
\includegraphics[width=0.7\textwidth]{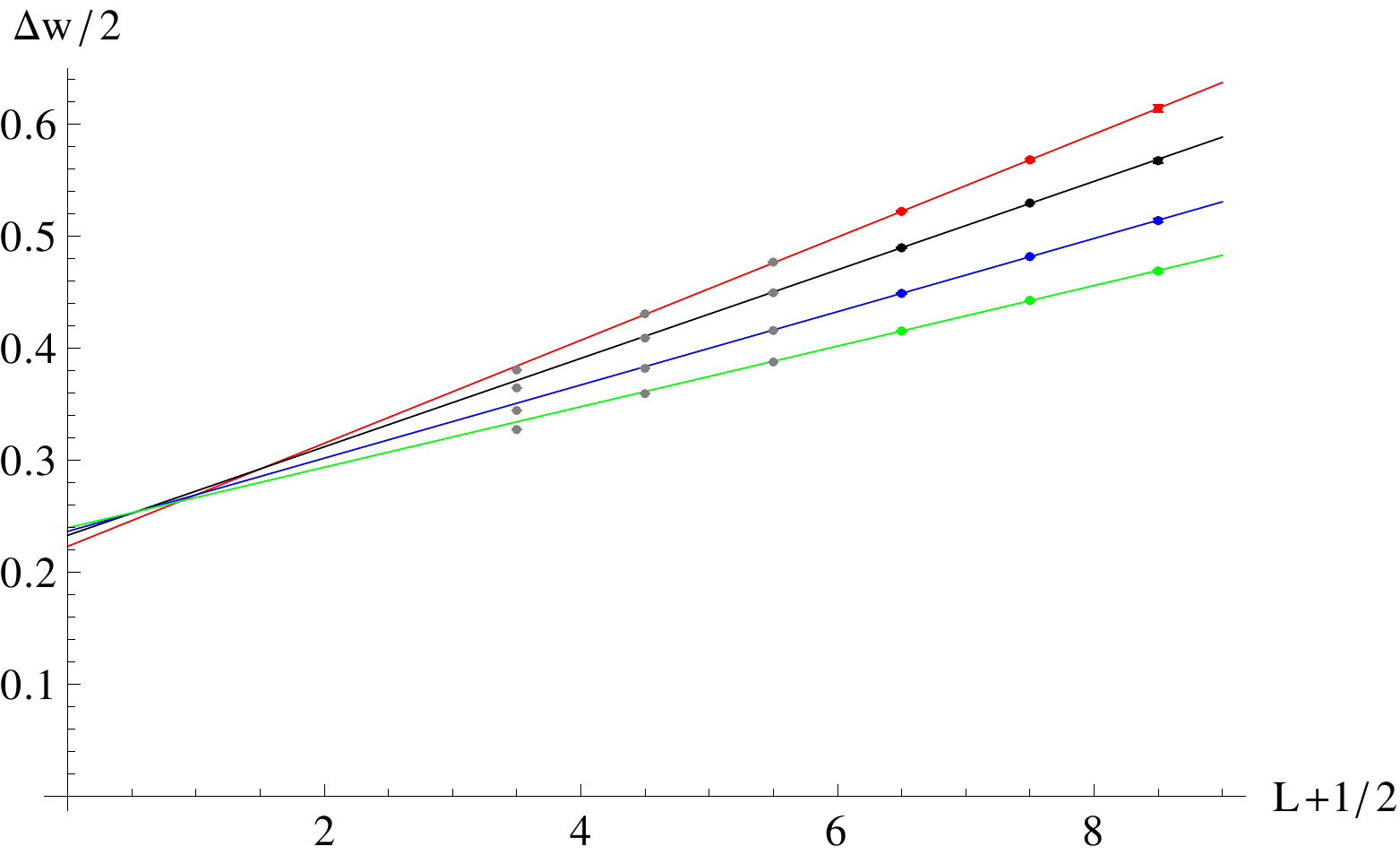}
\caption{Plots of $\frac{\Delta w}2=\frac 12 \left(w_\infty(L+1)-w_\infty(L)+\frac 12
  \log\left(1+\frac 1 L\right)\right)$ obtained with method 1) as a function
  of $L+\frac 12$ at $S=0.4$ and $b=0.36$ (red), 
  $b=0.362$ (black), $b=0.365$ (blue) and $b=0.368$ (green). Error bars are not visible in the plot. 
The straight lines show linear fits through the corresponding data points.
Only points $6 < L+\frac 12 < 9$ are used in the fits. The string tension values obtained 
from these fits are collected in~Table~\ref{tab:sigmas}.} 
\label{fig:SigmaExample} 
}

\FIGURE[htb]{
\includegraphics[width=0.9\textwidth]{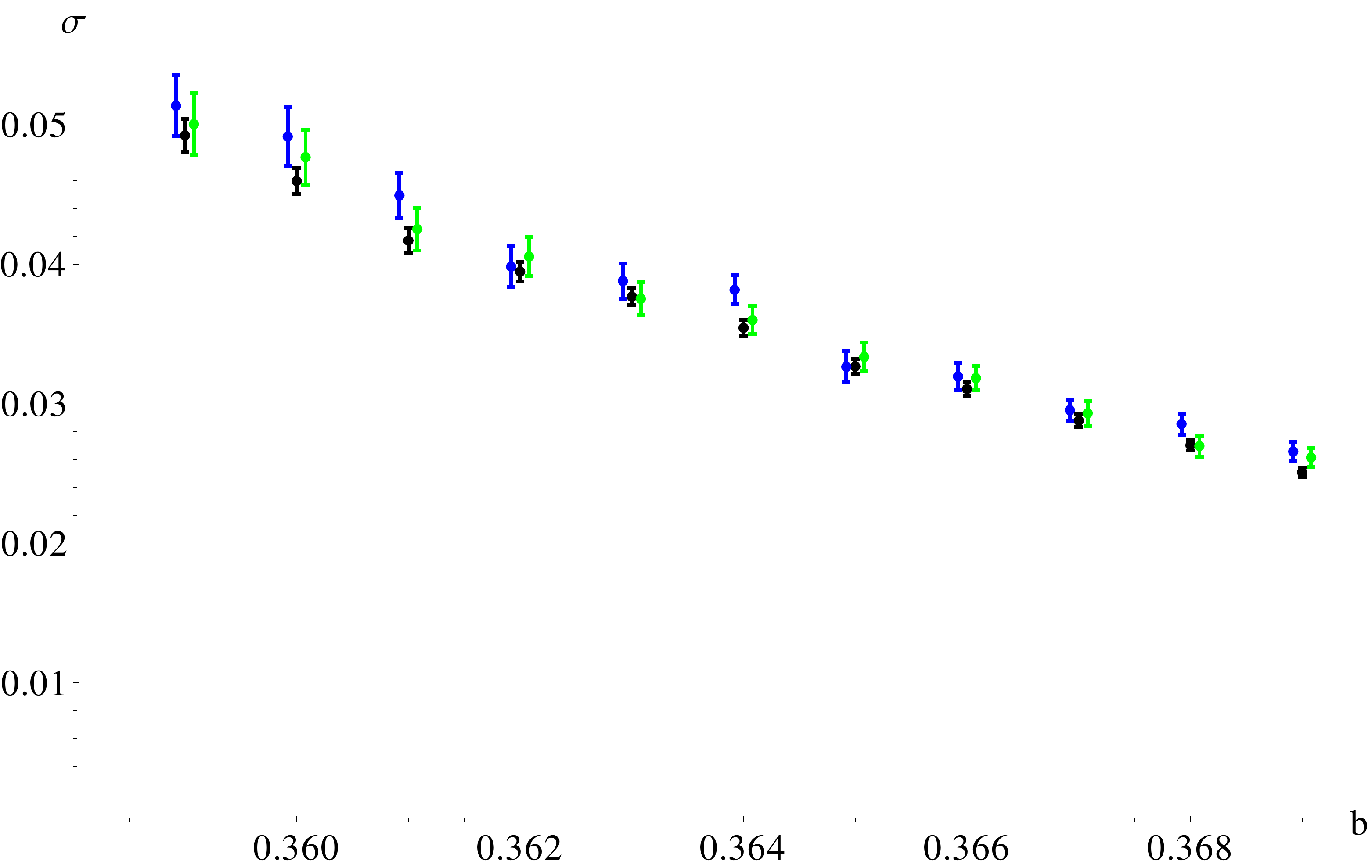}
\caption{Plots of the string tension values of 
  Table~\ref{tab:sigmas}: method 1) in black, method 2a) in green, method 2b) in
  blue. $b$-values for 2a) and 2b) are slightly shifted to opposite sides of the 
  true $b$-value for visibility.} 
\label{fig:Sigma-b} 
}

\subsection{Continuum limit}

A scale length in lattice 
units denoted by $\xi_c(b)$ is used to carry out extrapolations to the continuum
limit. It is defined at $N=\infty$ using a three-loop calculation of 
the $\beta$-function for the lattice coupling. 
The coefficients are written as ${\bar \beta_i}$, $i=0,1,2$. 
\begin{align}
\bar \beta_0&=\frac{\beta_0}{N}=\frac{11}{48 \pi^2}\,,\\
\bar \beta_1&=\frac{\beta_1}{N^2}=\frac{34}{3 (16 \pi^2)^2}\,,\\
\bar \beta_2&=\lim_{N\to\infty} \frac{\beta_2}{N^3}\approx -3.12211\times 10^{-5}\,.
\end{align}
Integrating the RG flow, we define:
\begin{align}\label{eq:Lc}
\xi_c(b)=0.26 \left(\frac{\bar \beta_1}{\bar \beta_0^2}+\frac{b_I(b)}{\bar
  \beta_0}\right)^{-\frac{\bar \beta_1}{2\bar\beta_0^2}}\, \exp\left[\frac{b_I(b)}{2\bar
  \beta_0}\right]\, \exp\left[\frac{\bar \beta_2}{2\bar\beta_0^2 b_I(b)}\right]\,.
\end{align}
Above we have replaced the gauge coupling $b$ by $b_I(b)$,
\begin{align}
b_I(b)=\lim_{N,V\to\infty} b\, W_N(L=1,b,S=0,V)\,,
\end{align}
 a substitution known
as tadpole improvement. In practice this substitution gets one much more rapidly into 
the asymptotic regime where truncated perturbation theory reasonably accurately
describes the RG flow. 
The improved infinite-$N$ coupling constant $b_I(b)$ is defined without smearing. 

The definition of $\xi_c(b)$ is taken to match with \cite{alton}. 
We only added a numerical prefactor to make  $\xi_c(b)\approx L_c(b)$, where $L_c(b)$ 
is given in~\cite{plb-red}. 
This approximation is good to 10-15\% in our range of couplings and would 
become exact at $b=\infty$. 
It is well known that direct continuum extrapolations 
are subject to large systematic errors since one is too far from a truly
asymptotic regime. There are many prescriptions for what to use. We
do not have a particular opinion about which one is best. Our objective in 
this paper is not to obtain the highest accuracy possible with our data. We preferred
to choose a reasonable existing prescription to facilitate comparison with other numerical work
on closely related topics. The net consequence of this is that numerical 
results for the continuum limit are substantially less accurate than numerical results at finite 
lattice spacing. The work of different groups may sometimes be compared only in the 
continuum limit, and sometimes also at finite lattice spacing. It is possible that one ends 
up with a statistically significant discrepancy at finite lattice spacing but no statistically 
significant discrepancy in the continuum limit. This will be the situation for our determination of the
string tension, but this remark has to be qualified by the 
comment that one needs to add the observation that 
on the lattice one should get identical numbers for the string tension extracted from the 
two point function of Polyakov loops and that extracted from rectangular loops.

We separately carry out
two two-parameter fits of the relation between the string tension $\sigma(b)$ 
and $\xi_c(b)$. The two pairs of parameters are denoted by 
$d_0$, $d_1$ and $f_0$, $f_1$:
\begin{align}\label{eq:continuum1}
\sigma(b)=\frac{d_0}{\xi_c(b)^2}+\frac{d_1}{\xi_c(b)^4}
\end{align}
and 
\begin{align}\label{eq:continuum2}
\frac1{\xi_c(b)^2}=f_0^{-1} \sigma(b)+f_1 \sigma(b)^2\,.
\end{align}
We use ranges $0.359 \leq b \leq 0.369$ (range A) and $0.362 \leq b \leq 0.367$ (range
B). We also use the limited $b$ range (B) since we have observed increasing
$\chi^2$'s for the infinite-$N,V$ extrapolations using method 2a) for $b\leq
0.361$. This was mentioned in~Sec.~\ref{Sec:NVlimits}. Another reason is that finite-volume
effects increase with increasing $b$.    This reason only applies to method 1). 

Results for $d_0$ and $f_0$ using the  $\sigma(b)$ values in 
Table~\ref{tab:sigmas} are given in Table~\ref{tab:ContRes}. 
Plots of the corresponding fit functions \eqref{eq:continuum1} 
and \eqref{eq:continuum2} are shown in Fig.~\ref{fig:contfit1}.

\FIGURE[htb]{
\includegraphics[width=0.9\textwidth]{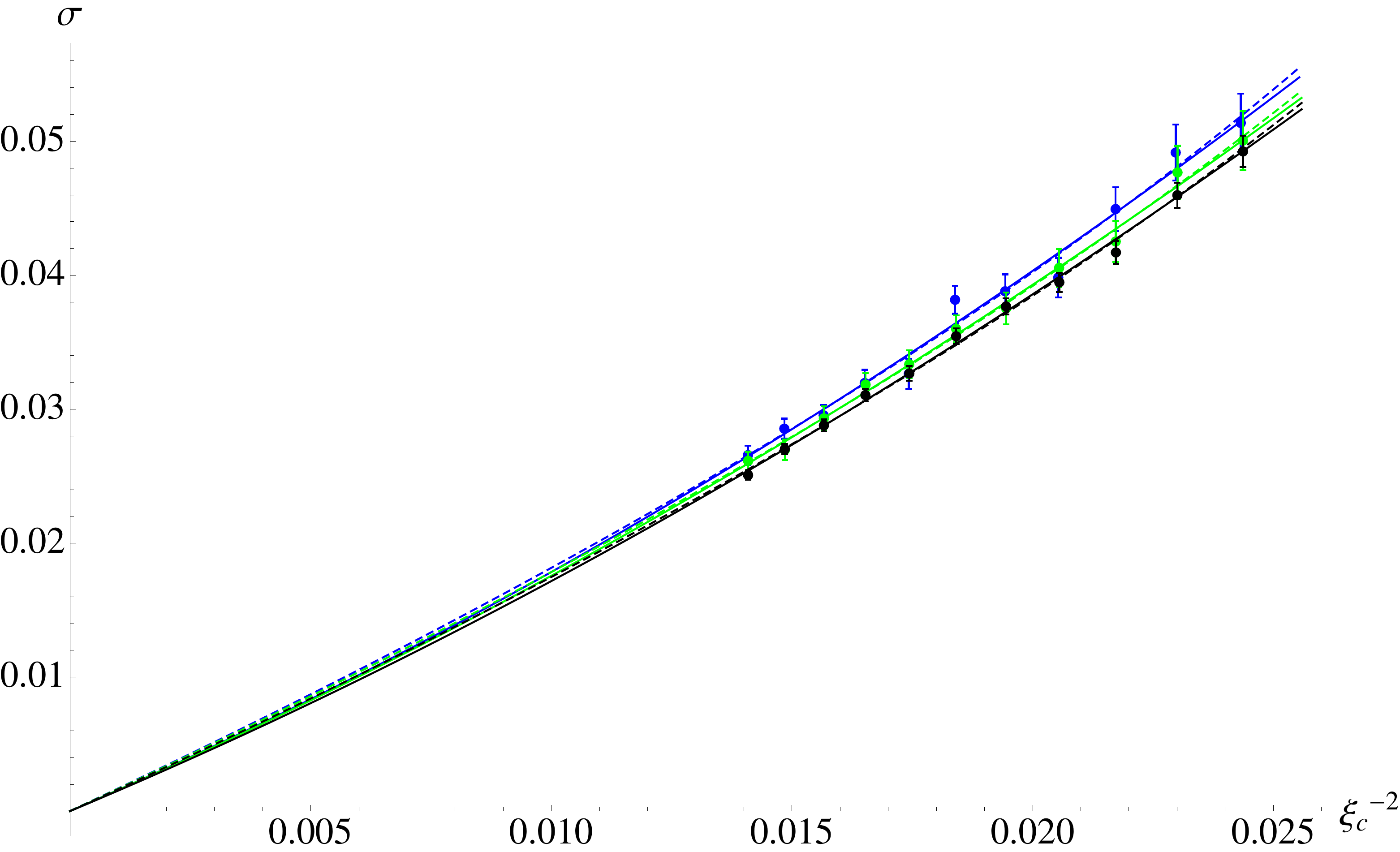}
\caption{Plots of $\sigma$ as a function of $\xi_c^{-2}$: 
method 1) in black, method 2a) in green, method 2b) in blue, 
together with fit functions \eqref{eq:continuum1} (solid lines) 
and \eqref{eq:continuum2} (dashed lines). 
The fits are obtained using $0.359\leq b \leq 0.369$ (range A); 
results for $d_0$ and $f_0$ are given in Table~\ref{tab:ContRes}.} 
\label{fig:contfit1} 
}

The difference between the two fits is a simple indicator of systematic errors
induced by the truncation of the perturbative series. 
For all continuum extrapolations $d_0<f_0$. These particular systematic deviations
are of the same order as the  statistical errors.

The result of Allton et al.~\cite{alton} is $\sigma \xi_c^2 \to 1.85$ in our notation 
with a statistical error of 1\% and a systematic error of 16\% on the 
continuum result. Their
numbers were extracted from Polyakov loops which are substantially
longer than one side of our rectangular loops. Apparently, 
the physical length of string that
came into their calculations allows for a substantially more accurate determination
of the string tension. 
The systematic errors are dominated by the extrapolation to continuum and 
their relative size is roughly the same for us. Although we work at somewhat higher $b$-values, 
this has little impact on the extrapolation. Our $b$'s are still too far from the full set-on
of perturbative asymptotics. 

Setting $\bar \beta_2=0$ in the expression for $\xi_c(b)$
(cf.~Eq.~\eqref{eq:Lc}) results in an increase of about 26\% for $d_0$, and an
increase of about 29\% for $f_0$, cf. Table~\ref{tab:ContResBeta2NULL} and Fig.~\ref{fig:contfit2}.
This number is too large to commit to a precise estimate of the systematical error. 
We could be optimistic and assume that the next term in perturbation theory
would make a smaller correction but it is hard to tell. 

\FIGURE[htb]{
\includegraphics[width=0.9\textwidth]{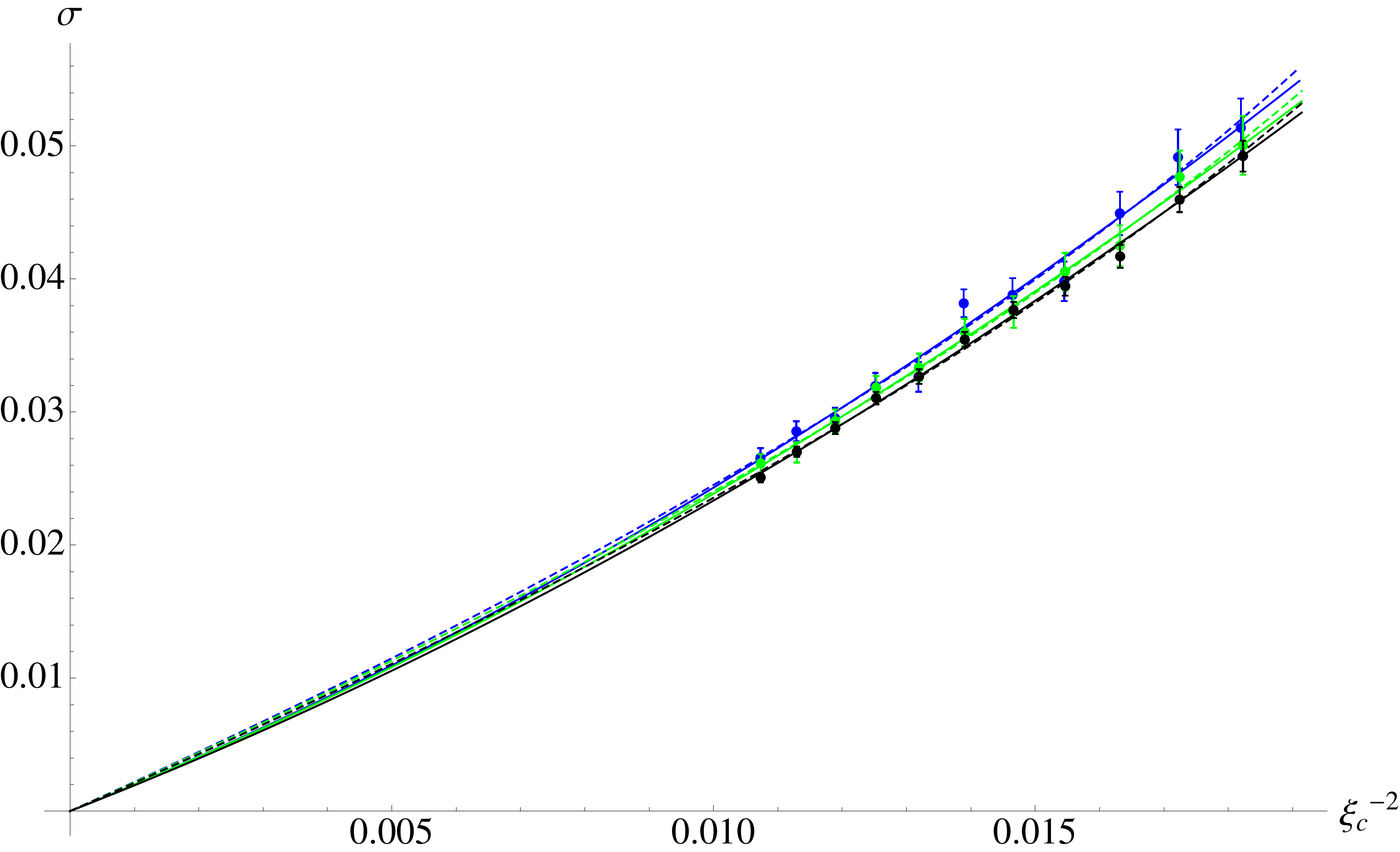}
\caption{Same as Fig.~\ref{fig:contfit1} for $\bar \beta_2=0$ in $\xi_c(b)$ 
(cf.~Eq.~\eqref{eq:Lc}). Fit results are given in Table~\ref{tab:ContResBeta2NULL}.} 
\label{fig:contfit2} 
}

Later we shall find some rough estimates of the effective coupling constant 
$\frac{g^2 N}{4\pi}$ at our smearing levels and it would come out to be of order
unity. The definition of $\xi_c(b)$ is at zero smearing, so these effective couplings are 
not directly relevant. Nonetheless, the large systematic error coming from truncating perturbation
theory comes as no surprise. 

In summary, we find that the infinite-$N$ continuum string tension 
is given by
\begin{align}
\lim_{b\to\infty}\sigma(b)\xi^2_c(b)=1.6(1)(3)\,.
\end{align}
The first error is statistical and the second systematic. The systematic error is 
more of a guess than a well founded estimate. 
Our central number is 2-3 standard deviations smaller than that of~\cite{alton}. 
In terms of $\Lambda_{\overline{MS}}$, this translates to:
\begin{align}
\sigma/\Lambda_{\overline{MS}}^2=3.4(2)(6)\,.
\end{align}
A previous estimate for the string tension at infinite $N$ extracted from rectangular 
loops has been given in \cite{narkis}. Expressed in terms  of our variables it
is $\sigma\xi_c^2\vert_{b_I=0.182} =2.2(3)$. 
The results of \cite{narkis} are
claimed to be compatible with~\cite{alton} at infinite $N$.
These results were obtained working at $N=37,47,59$ on a $6^4$ lattice at small 
values $b=0.345,0.348,0.350$ assuming large-$N$ reduction and 
including folded loops in the analysis. 

While writing up this paper a new study~\cite{tony} appeared which also deals with
rectangular Wilson loops with the objective to test a new method of full 
twisted large-$N$ reduction~\cite{tony-prev}. 
This successful test is carried out on the physical value of 
the string tension. These authors obtain
$\frac{\sigma}{\Lambda_{{\overline MS}^2}} = 3.63(3)$ (statistical error) 
at $N=\infty$ in the 
continuum if they apply the continuum extrapolation method of~\cite{alton}. 
This number is fully consistent with ours and has very small errors by
comparison. The number in~\cite{alton} 
is $\frac{\sigma}{\Lambda_{{\overline MS}^2}} = 3.95(3)(64)$ at $N=\infty$ in the 
continuum. 

There seems to be a disagreement at the statistical level between~\cite{alton} and 
our result which agrees with that of~\cite{tony}. The result of~\cite{narkis} seems to
side with that of~\cite{alton}, but has too large errors to be sure. 
The systematic error is too big to claim evidence for a difference 
between the string tension extracted from Wilson loops and that extracted from
Polyakov loop correlators by~\cite{alton} in the continuum limit. 
Such a difference would be very difficult to accept at the theoretical 
level. Given the differences between these various simulations the case for a real 
discrepancy between the Wilson loop string tension and the Polyakov correlator
string tension at the lattice level is not worrying so far. If one ignores smearing, 
these two string tensions ought to be 
equal already at the lattice level. It might be of interest to settle this in future work.

\subsection[String tension at finite $N$]{\boldmath String tension at finite $N$}
\label{sec:SigmaFinN}

We now determine the string tension in lattice units at finite $N$.
Extrapolating to infinite $N$ this would show how the string tension in 
lattice units converges to its infinite-$N$ limit. We do this in order to
get a feel about the commutativity of the limits $N\to\infty$ and
UV-cutoff $\to\infty$. 

We determine the string tension $\sigma_N$ at fixed $b$, $S$, $V$ and $N$, 
by fitting $c_{2,N}$ and $\sigma_N$ in  
\begin{align}\label{eq:SigmaFromSquaresFinN}
\frac 12 \left(w_N(L+1)-w_N(L)+\frac 12 \log\left(1+
\frac 1 L\right)\right)=\sigma_N \left(L+\frac 12\right)+\frac{c_{2,N}}{2}
\end{align}
to $6\leq L \leq 9$ square loop data. 
We use data obtained on volumes $V=24^4$, $18^4$, $14^4$, $12^4$ 
for $N=7$, $11$, $19$, $29$, respectively, as detailed earlier. We already mentioned in Sec.~\ref{Sec:NVlimits} that
for these cases we believe that finite-volume effects are negligible.
For $N=7$, $11$, $19$ we use $S=0.52$, and for $N=29$ we use $S=0.4$. 
As before, there is no dependence on $S$ within
statistical errors and the errors decrease with increasing $S$ 
(cf.~Sec.~\ref{sec:Sdep} below).  

Figure~\ref{fig:SigmaFinN} shows plots of the finite-$N$ string tensions as a
function of $b$ together with the infinite-$N$ result obtained from
$w_\infty$ by method 1) in Sec.~\ref{sec:infNSigma}. 
For fixed $b\leq 0.367$ the results for the string tension at finite $N$ are well
described by
\begin{align}\label{eq:sigmaFinNCorr}
\sigma_N(b)=\sigma_\infty(b)+\frac{h(b)}{N^2}\,.
\end{align}
Results for $\sigma_N$ and fit parameters $\sigma_\infty$, $h$ are given in
Table~\ref{tab:sigmaN}. The infinite-$N$ string tension obtained in this
manner is in good agreement with our previous results in~Table~\ref{tab:sigmas}.
Those results were determined from $w_\infty$. 
We see that $h(b)\sim 10 \sigma_\infty (b)$. A 
relative variation of order $\frac{10}{N^2}$ may seem large. 
At fixed lattice coupling $b$ 
one would need to go to $N$ values of 
order 20 to be able to credibly extrapolate
to infinite $N$. 

It will become clear later on, in section~\ref{sec:cont_lim_fin_N}, 
that the large coefficient of
the $\frac{1}{N^2}$ term is replaced by a much smaller one when one looks at $\sigma$ 
not as a function of $b$, but as a function of $b$ times the plaquette average. The
plaquette average also has a finite $N$ correction and the fact that it works in the manner
described is relatively well know, as mentioned in Sec.~\ref{sec:cont_lim_fin_N}.

\FIGURE[htb]{
\includegraphics[width=0.9\textwidth]{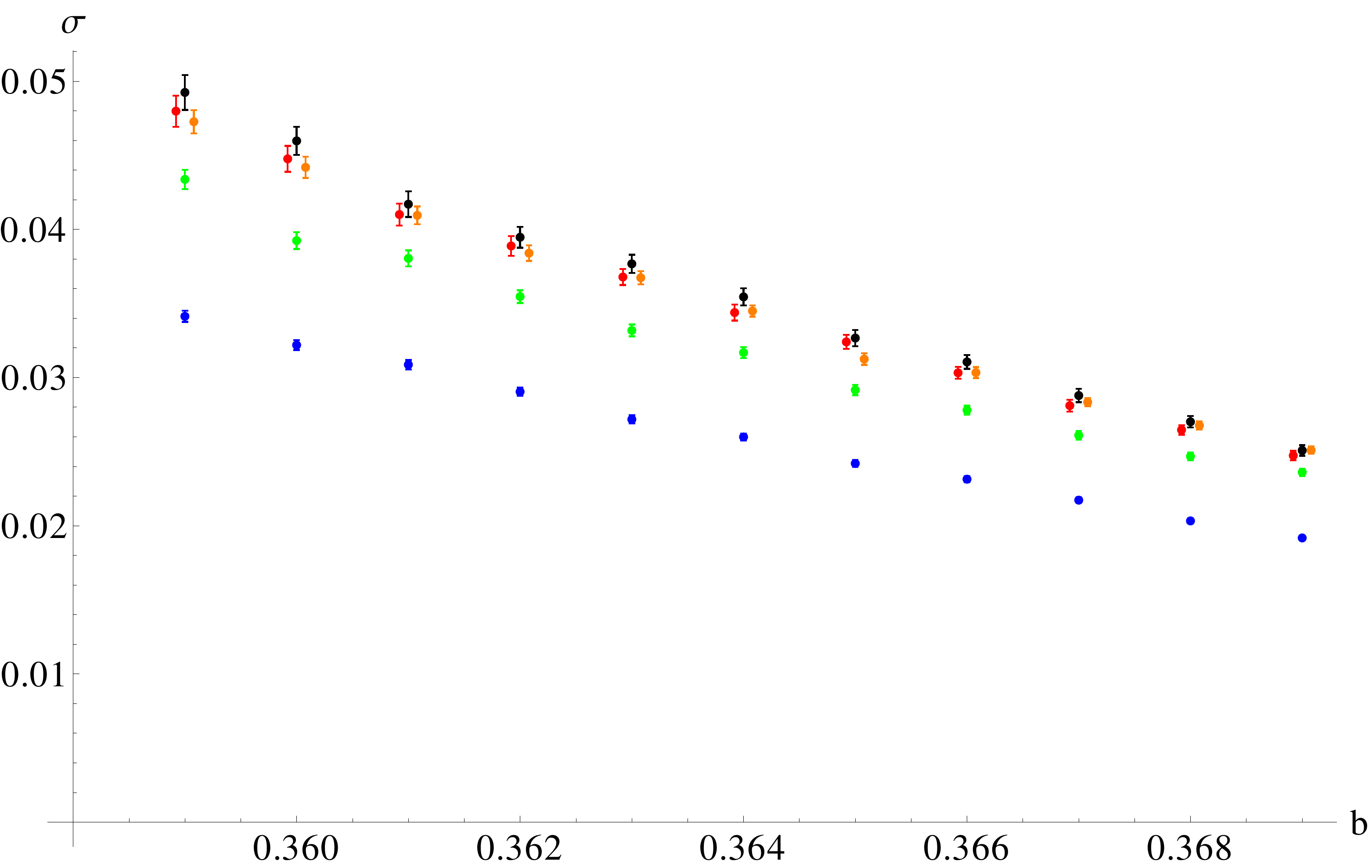}
\caption{Plots of the finite-$N$ string
  tension $\sigma_N (b)$ from square loops for $N=7$ (blue), $N=11$ (green),
  $N=19$ (orange), $N=29$ (red) together with the infinite-$N$ result (black) obtained by
  method 1), cf.~Table~\ref{tab:sigmas}. $b$-values for $N=19$ 
  and $N=29$ are slightly shifted in $b$ in the plot for visibility.} 
\label{fig:SigmaFinN} 
}

\subsection{Smearing dependence}
\label{sec:Sdep}

Figures \ref{fig:budgetS14} and \ref{fig:budgetS26}  show plots of $w_N$ as a
function of $L$ for two different smearing levels $S=0.28$ and $S=0.52$, for
$N=11$, $V=18^4$, $b=0.365$. Also shown are plots of the
individual terms $w_N$ is composed of (cf.~Eq.~\eqref{eq:wL1L2}). 
The variation in $\sigma_N$ is about 1.9\%, of the order of the statistical errors
(1.2\% for $S=0.52$, 2.6\% for $S=0.28$). $c_2$ and $c_1$ exhibit larger variation. 

These figures allow us to see the relative sizes of the various contributions 
to the Wilson loop at fixed smearing and how these relative
contributions change when smearing is changed. For a small amount
of smearing the negative of the logarithm of the 
Wilson loop is larger (see the red data 
points and the total fit drawn as a black curve). This mainly reflects the 
larger size of the perimeter term whose contribution is larger than that of
the area term. The log-term also makes a sizable contribution, smaller than
that of the area term. The smallest contribution comes from the shape-dependent
constant. The logarithmic dependence on smearing of the latter has small
numerical impact. 

Wilson loops
larger than their critical size smoothly merge with their behavior for 
small sizes. One can imagine how all terms except the area one
do this. The area term has to morph into something else. A likely 
candidate is a term going as the area squared (for planar loops). This term 
comes from the dimension four condensate which would enter in 
an expansion in loop size small relative to the extent of smearing. 
For a square loop of side $l$ and smearing $s$, one gets from one gluon 
exchange $w_N=\frac{g^2 N}{4\pi} \frac{1}{128\pi} \frac{l^4}{s^2}$.  
So, we only need something like $l^2/s < 20 $ for this approximation to 
become relevant.

\FIGURE[htb]{
\includegraphics[width=0.7\textwidth]{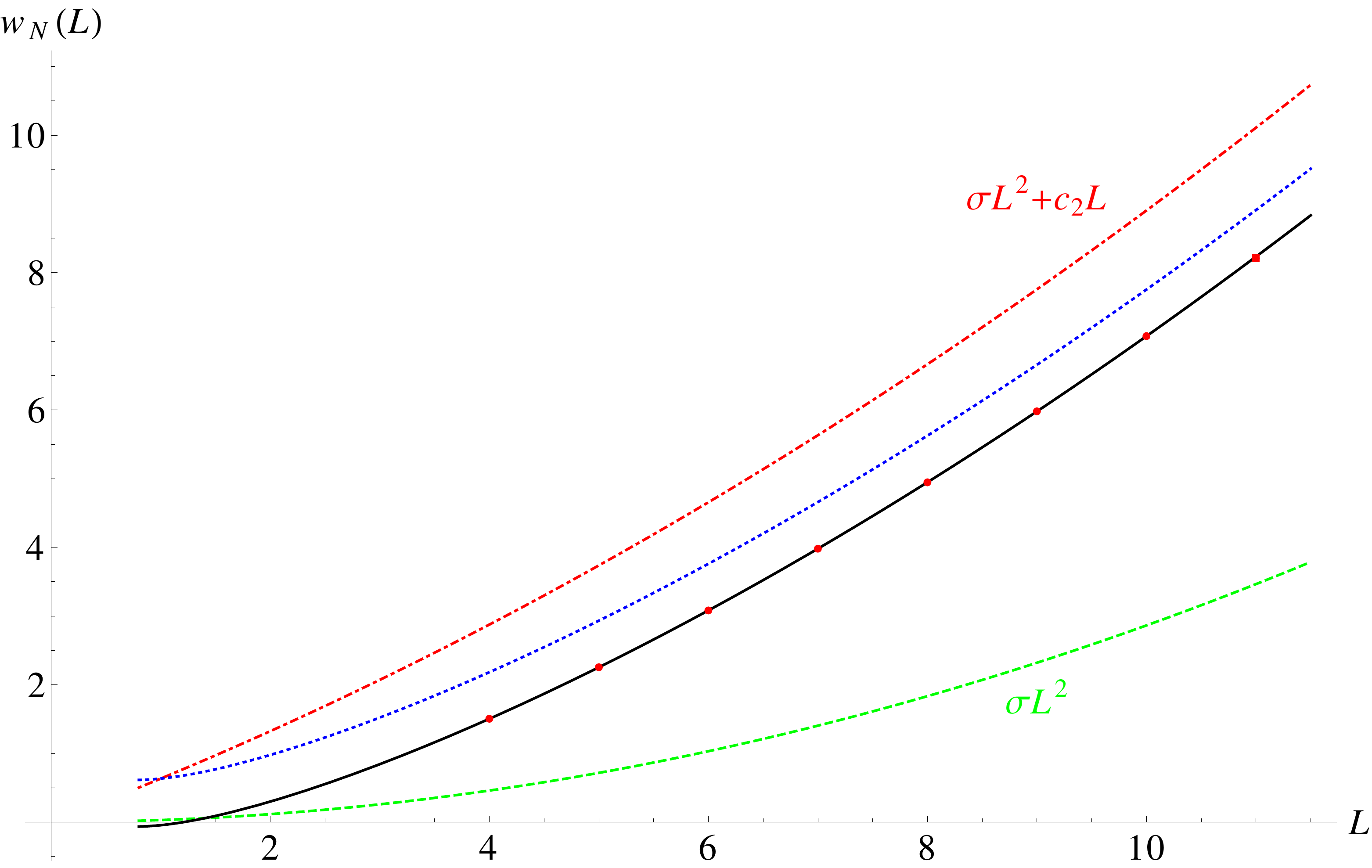}
\caption{Plot of $w_N(L)$ for $N=11$ on $V=18^4$ at $b=0.365$ and $S=0.28$
  (red points, error bars are not visible), together with analytic
  functions $\sigma L^2$ (green, dashed), $\sigma L^2 + c_2 L$ (red,
  dotdashed), $\sigma L^2+c_2 L-\frac 12 \log L$ (blue, dotted), and $\sigma
  L^2+c_2 L-\frac 12 \log L+c_1$ (black, solid). The fit parameters used to
plot the analytic functions were obtained from the data 
at $6\leq L \leq 9$. They are $\sigma=0.02863$, $c_2=0.6041$, $c_1=-0.6788$.} 
\label{fig:budgetS14} 
}

\FIGURE[htb]{
\includegraphics[width=0.7\textwidth]{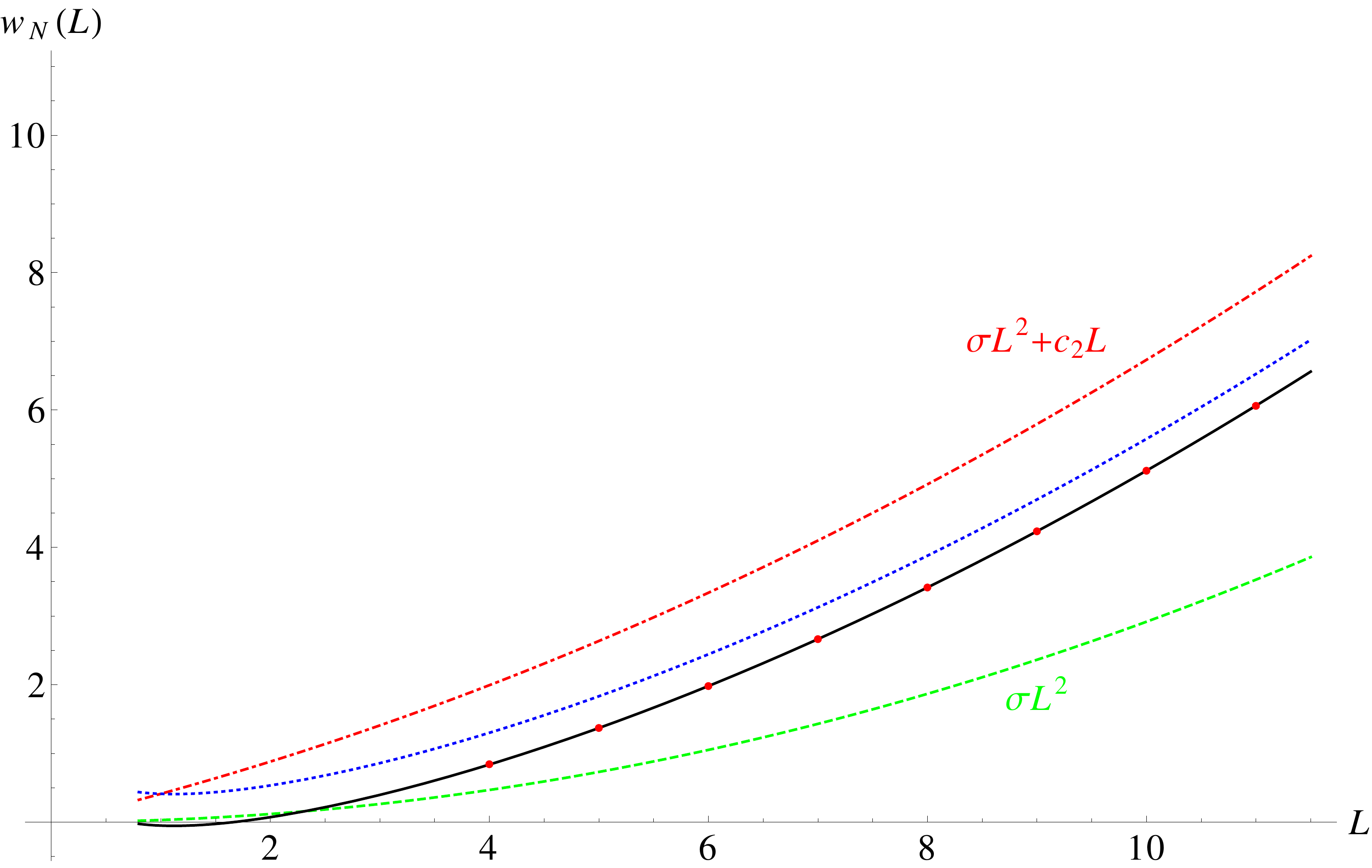}
\caption{Same as Fig.~\ref{fig:budgetS14} for $S=0.52$. Here,  $\sigma=0.02916$, $c_2=0.3813$, $c_1=-0.4624$.}  
\label{fig:budgetS26} 
}

The level of smearing in lattice units was varied in a large enough range to contain 
levels of smearing in physical scales that are relevant both to the proximity of the 
large-$N$ transition and to testing whether there is a dependence on smearing in 
the continuum limit. 
We see that the string tension does not depend on the smearing
parameter within statistical errors. These errors decrease rapidly with increasing $S$
(cf.~Fig.~\ref{fig:SigmaSDepN11} for an example). 

We can check whether the expected continuum divergences as $s\to 0$ indeed 
are detected. On the lattice there will be no divergence as $S\to 0$, since
the lattice spacing regulates all divergences, regardless of 
whether they come from the Lagrangian
or from the observable. With a reasonable amount of smearing one detects a window
where the amount of smearing exceeds the lattice spacing influence but is still small
enough to exhibit the behavior that would have caused a divergence in the continuum.
Fig.~\ref{fig:c2SDepN11} shows 
that the $c_2$ coefficient of the perimeter term increases
linearly with $\frac1{\sqrt S}$, $c_2=c_2^{(0)}+c_2^{(1)}/\sqrt{S}$ in this window. 
Similarly, Fig.~\ref{fig:c1SDepN11} shows 
that the $S$-dependence of the $L$-independent term $c_1$ 
is consistent with a $\log(S)$ $S\to 0$ divergence: $c_1=-0.2538+0.3278 \log S$.

\FIGURE[htb]{
\includegraphics[width=0.7\textwidth]{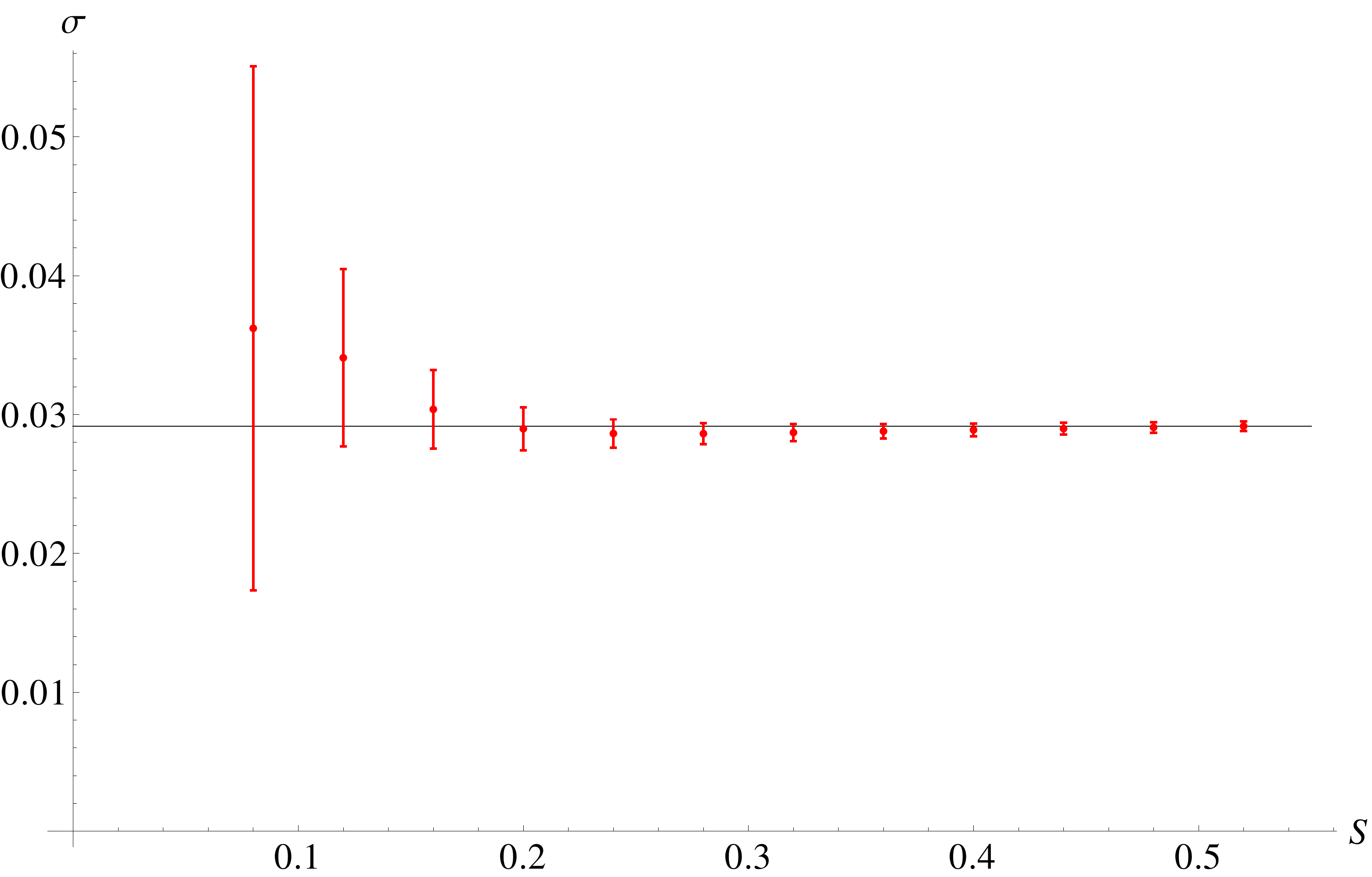}
\caption{String tension as a function of the smearing parameter $S$ for $N=11$,
  $b=0.365$, $V=18^4$. $\sigma$ is determined using square loops with $6\leq L
\leq 9$. The horizontal black line corresponds to $\sigma(S=0.52)=0.02916$.} 
\label{fig:SigmaSDepN11} 
}

\FIGURE[htb]{
\includegraphics[width=0.7\textwidth]{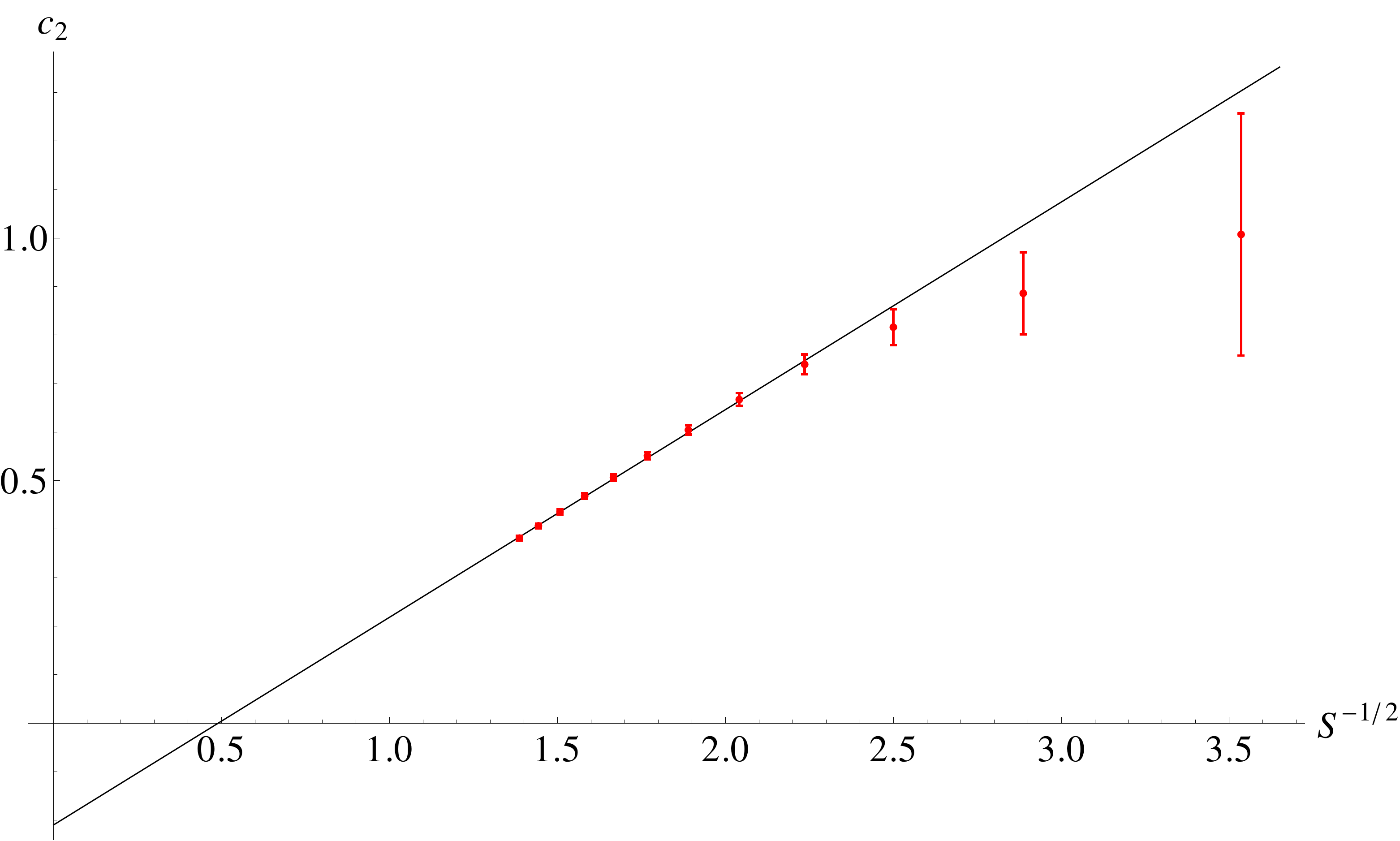}
\caption{Perimeter coefficient $c_2$ (determined from square loops with $6\leq L
\leq 9$) as a function of the smearing parameter $S$ for $N=11$,  $b=0.365$,
$V=18^4$. The straight black line shows the fit $c_2=  -0.2097 +
0.4279 /\sqrt S$.} 
\label{fig:c2SDepN11} 
}

\FIGURE[htb]{
\includegraphics[width=0.7\textwidth]{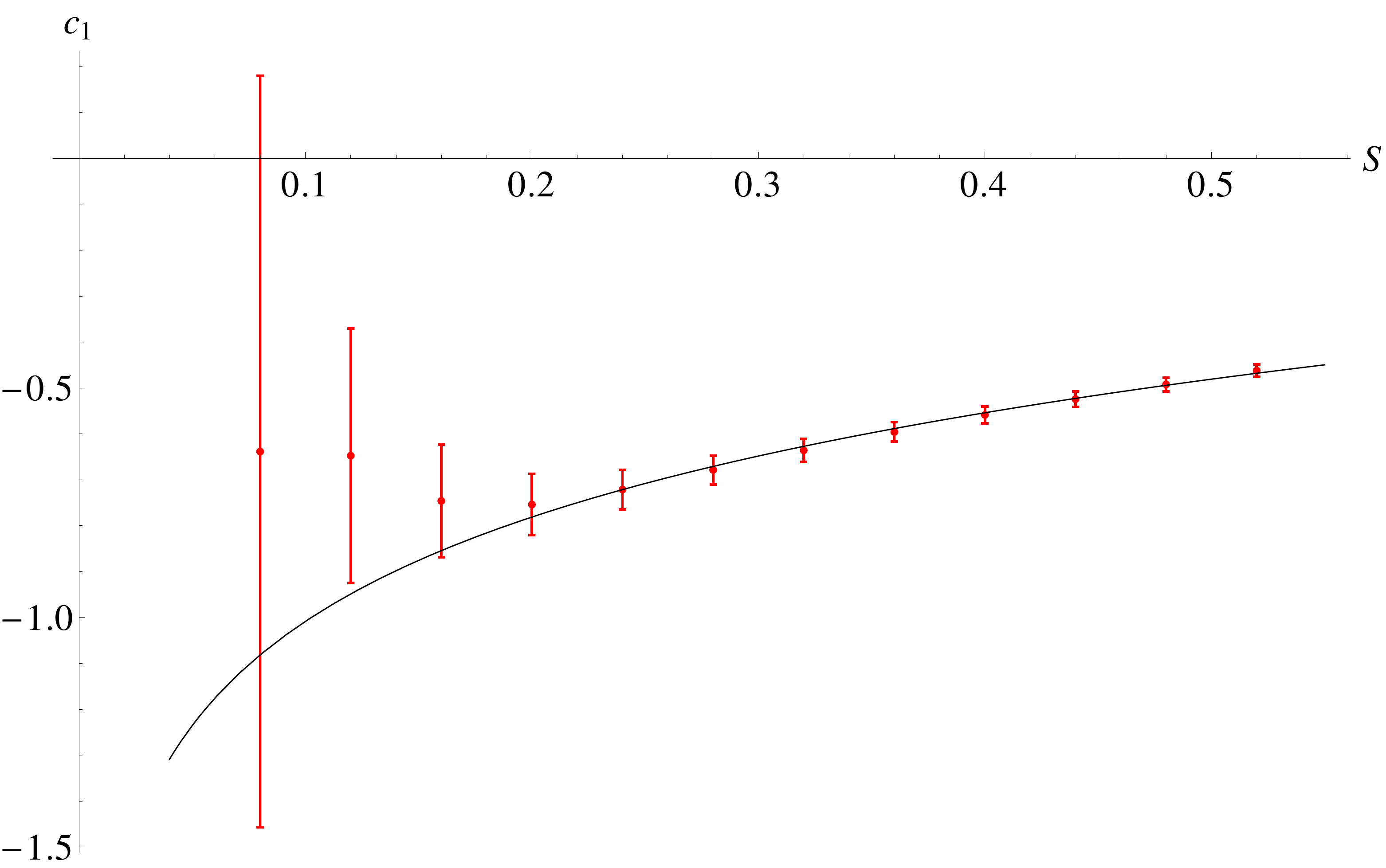}
\caption{$L$-independent constant $c_1$ (determined using square loops with
  $6\leq L \leq 9$) as a function of the smearing parameter $S$ for $N=11$,  $b=0.365$,
$V=18^4$. The black line shows the fit $c_1=-0.2538 +
0.3278 \log S$.} 
\label{fig:c1SDepN11} 
}

\subsection[Continuum limit for finite $N$]{\boldmath Continuum limit for finite $N$}
\label{sec:cont_lim_fin_N}
Most of the relatively large 
finite-$N$ corrections to the string tension get absorbed 
when considered at fixed improved coupling $b_I(b,N)$ rather than
fixed $b$. Figure~\ref{fig:SigmaFinNbI} makes this clear. $b_I(b,N)$ is determined by the unsmeared 
plaquette averages at the corresponding finite value of $N$. 
Tadpole improvement 
simultaneously improves the approach to continuum and to infinite $N$. 
Looking at $\sigma_N (b_I)$ instead of $\sigma_N(b)$ gives a better 
indication for the speed of large-$N$ convergence in the continuum. 

We extrapolate to the continuum at finite $N$ using
Eqs.~(\ref{eq:continuum1}, \ref{eq:continuum2}) with the finite-$N$
values for $\xi_c$, $\beta_2$ and $b_I$. Once we employ the improved
coupling scheme, we no longer observe a significant $N$-dependence in
$d_0(N)$ or $f_0(N)$. This is shown in Fig.~\ref{fig:d0f0}.  This
feature of tadpole improvement~\cite{alton} was first seen in the
context of Polyakov loop correlators and is reviewed
in~\cite{teper-zakopane}.

\FIGURE[htb]{
  \includegraphics[width=0.9\textwidth]{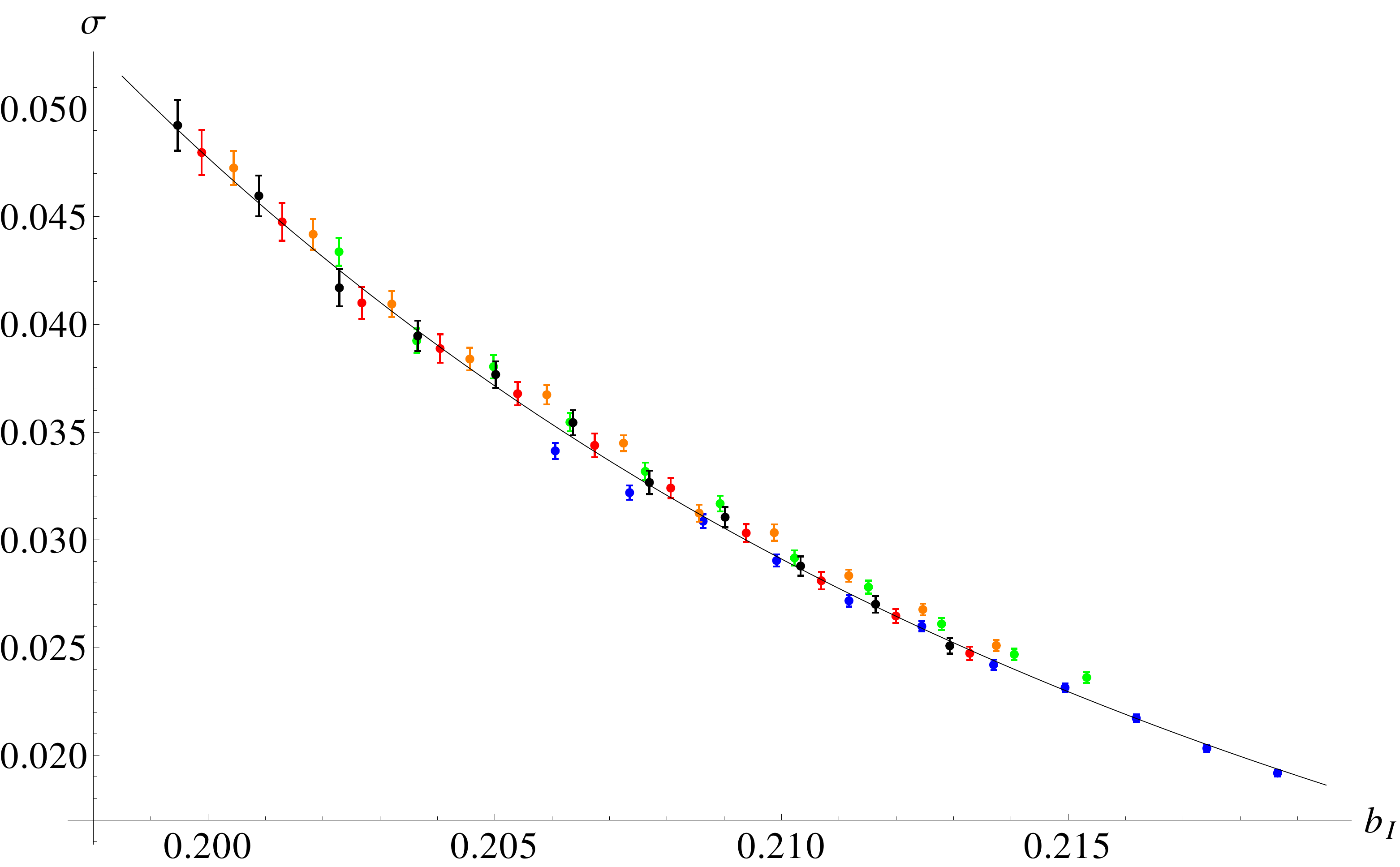}
  \caption{Same data as in Figure~\ref{fig:SigmaFinN} now plotted as a
    function of the improved coupling $b_I$. The solid line shows the
    result of the continuum extrapolation [method 1) \& range A in
    Table~\ref{tab:ContRes}] for the infinite-$N$ string tension,
    $\sigma(b)=1.50/\xi_c(b)^2+21/\xi_c(b)^4$,
    cf.~Eq.~\eqref{eq:continuum1}.}
  \label{fig:SigmaFinNbI}
}

\FIGURE[htb]{
  \includegraphics[width=0.7\textwidth]{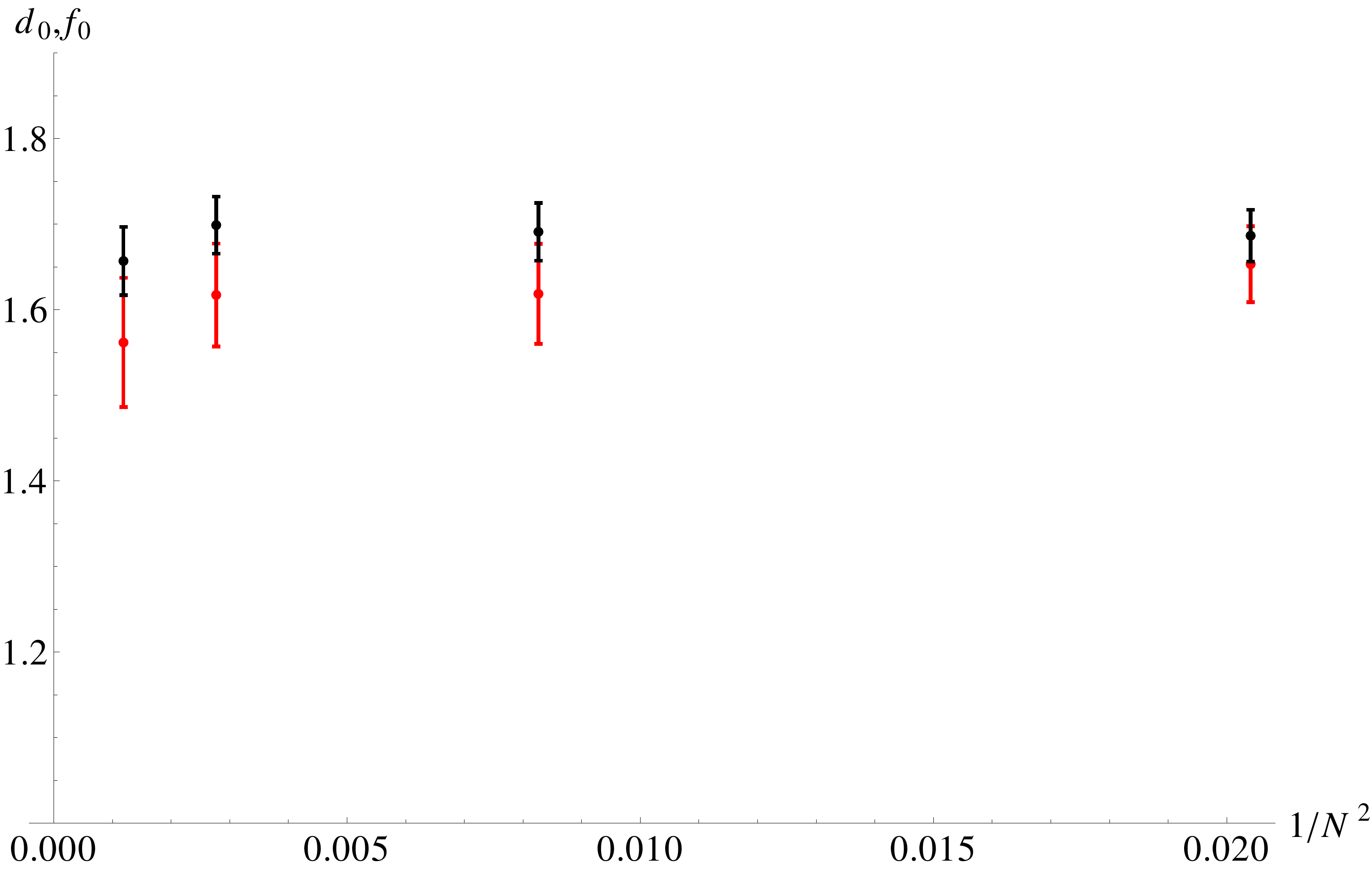}
  \caption{Plot of $d_0(N)$ (red) and $f_0(N)$ (black) determined from
    finite-$N$ versions of Eqs.~(\ref{eq:continuum1},
    \ref{eq:continuum2}) fitted in the range $0.359\leq b\leq 0.367$.}
  \label{fig:d0f0}
}

\subsection{Tree-level continuum perturbation theory}
\label{sec:PT}

We calculated 
the expectation value of a rectangular $l_1\times l_2$
loop in tree-level continuum perturbation theory. 
Diagrammatically, this gives a one-loop integral with a tree-level
smeared propagator. By ``tree-level" we refer to the absence of 
propagator and vertex radiative corrections.

We parameterize a general closed 
curve $\C$ by $x(u)$ with $x(0)=x(1)$ and $\dot x(u)\equiv\partial_u x(u)$).
To leading order in $g^2$,  $w_N^{\text{PT}}=-\log W_N$ is given by
\begin{align}
w_N^{\text{PT}}(\C,s)&=\frac 12 g^2 C_2 \int_0^1 du_1 \int_0^1du_2 \,D(x(u_1)-x(u_2);s) \delta_{\mu\nu}\dot x_\mu(u_1)\dot x_\nu(u_2)\,,\\
D(x;s)&=\int \frac{d^4 p}{(2 \pi)^4} \,e^{ip\cdot x} \frac 1{p^2} \,e^{- 2 s  p^2}\,.
\end{align}  
$C_2$ is the quadratic Casimir invariant given by $C_2^{\text{fund}}=\frac{N^2-1}{2N}$ 
in the fundamental representation. 
We first take a derivative w.r.t.~$s$ to bring the integrand into Gaussian form, 
\begin{align}
\partial_s D(x;s)=-\frac{1}{2^5 \pi^2 s^2} e^{-\frac{x^2}{8s}}\,.
\end{align}

Integrating around a rectangular $l_1\times l_2$ loop over $u_{1,2}$ then produces error functions. 
Next, we integrate over $s$, using $D(x,s)\to 0$ as $s\to \infty$.
\begin{align}
w_N^{\text{PT}}(l_1,l_2,s)=-\int_s^\infty ds^\prime\,
\partial_{s^\prime} w_N^{\text{PT}}(l_1,l_2,s^\prime)\, 
\end{align}  
leads to
\begin{align}\label{eq:wNPT}
w_N^{\text{PT}}(l_1,l_2,s)= \frac{g^2 C_2}{2}\, h\left(\frac{s}{l_1^2},\frac{s}{l_2^2}\right)
\end{align} 
with
\begin{align}
h\left(\frac{s}{l_1^2},\frac{s}{l_2^2}\right)=\frac{2}{\pi^2}\int_0^{\frac{l_1}{\sqrt{8s}}} dz \left(1-e^{-(l_2/l_1)^2 z^{2}}\right) \left(\frac{\sqrt{\pi}}2 \erf(z)+\frac1{2z}\left(e^{-z^2}-1\right)\right)+l_1\leftrightarrow l_2 \,.
\end{align}

For $s/l_i^2$ small, we get 
\begin{align}\label{eq:hPT}
h\left(\frac{s}{l_1^2},\frac{s}{l_2^2}\right)=\frac{1}{(2\pi)^{\frac 32}}\left( \frac{l_1+l_2}{\sqrt{s}}\right)+\frac
1{\pi^2} \log\left(\frac{s}{l_1l_2}\right)+h_0\left(\frac{l_2}{l_1}\right)+\mathcal O\left(\frac{s}{l_i^2}\right)\,.
\end{align} 
$h_0(l_2/l_1)=h_0(l_1/l_2)$ has the following integral representation:
\begin{align}
&h_0\left(\frac{l_2}{l_1}\right)=\frac{2}{\pi^2}\int_0^{1} dz \left(2-e^{-(l_2/l_1)^2 z^{2}}-e^{-(l_1/l_2)^2 z^{2}}\right) \left(\frac{\sqrt{\pi}}2 \erf(z)+\frac1{2z}\left(e^{-z^2}-1\right)\right)
\nonumber\\
&\quad+\frac{2}{\pi^2}\int_1^{\infty} dz\left[
\left(2-e^{-(l_2/l_1)^2 z^{2}}-e^{-(l_1/l_2)^2 z^{2}}\right) \left(
\frac{\sqrt{\pi}}2\erf(z)+\frac1{2z}\left(e^{-z^2}-1\right)\right)-\sqrt{\pi}+\frac{1}{z}\right]\nonumber\\
&\quad+\frac{3}{\pi^2}\log2-\frac2{\pi^{\frac32}}\,.
\end{align}

Beyond the perimeter divergence a loop with kinks will
have logarithmic singularities as $s\to 0$. At each kink 
we denote by $\gamma_i$ the angle between the tangents. 
The well known expression is:
\begin{align}
\label{eq:ptcorner}
\log(W_N)_{\rm corners} =-\sum_i
\frac{g^2 C_2({\rm fundamental})}{4\pi^2} (\gamma_i \cot\gamma_i-1 )
\log\left ( \frac{{\rm Length}(C)}{\sqrt{s}}\right )\,.
\end{align}
Exactly backtracking segments of $\C$ should cancel out for any $s\ne 0$. 
Nonetheless, the coefficient of the $\log(s)$ divergence above 
blows up as $\gamma_i\to\pi$. 
The limits $s\to 0$ and $\gamma_i\to \pi$ do not commute. 
This is not a surprise, because when a finite loop segment becomes exactly backtracking 
the perimeter changes discontinuously, affecting already the leading term in the $s\to 0$
asymptotic series. 

Note the appearance of the logarithm of the perimeter at leading order 
in perturbation theory. There is a well known 
fundamental difference between the perimeter divergence and the logarithmic 
one in higher orders of perturbation theory~\cite{corner-div}: While the 
perimeter divergence maintains its tree-level dependence on the loop to all orders, 
the corner divergence does not because it corresponds to a kink-angle 
dependent anomalous dimension. 
Were we to choose to define the overall scale as the square root of the minimal area, this
term would add a shape dependence to the logarithm of the Wilson loop. 
The logarithm of the perimeter comes in from the integral over gluon exchanges where 
the gluon connects point on the opposing sides of the corner. When both endpoints 
are close to the corner the propagator is conformal approximately and this generates
a logarithm of the distance along each side of the angle. For a very dilated loop 
having a finite number of separated kinks, one expects to leave the conformal regime
before any other kink is encountered. This would replace the logarithm of the 
perimeter in the corner divergence term by $\log\frac{1}{\Lambda}$. 
We assumed that $s$ stays fixed, of the order $\frac{1}{\Lambda^2}$.
As we mentioned already, there are corner terms going as $\log^\kappa s$ 
at higher order which sum up to $\log(\log s)$ in the LLA. 

\subsubsection{Perimeter coefficient}

We have seen numerically that there is a term in the logarithm of 
the smeared Wilson loops which is proportional to the perimeter
and diverging as $s^{-\frac 12}$ for $s\to 0$. This is a local
divergence on the loop and therefore ought to be 
calculable in perturbation
theory. One expects no $\log(s)$ divergent contributions to the perimeter term at 
all orders in perturbation theory. 

From~Eqs.~(\ref{eq:wNPT}, \ref{eq:hPT})
we get the following tree-level formula for the perimeter term 
 for a square $l\times l$ loop:
\begin{align}
\frac{g^2 C_2}{2}\,\frac{1}{(2 \pi)^{\frac 32}}\, \frac{ 2l}{\sqrt{s}}\,.
\end{align}

For example, at $b=0.365$ and $N=11$, we have obtained
$c_2=-0.2097+0.4279/\sqrt{S}$ (cf.~Fig.~\ref{fig:c2SDepN11}). Matching the
coefficient of the $L/\sqrt{S}$ term with tree-level PT would require
$\frac{g^2 N}{4\pi}\approx 1.08$. This is an indication for the order of magnitude for
an effective running coupling at this level of smearing.

\subsubsection[Coefficient of $\log(s)$]{Coefficient of \boldmath $\log(s)$}
\label{sec:coeff-log-s}
The corner divergence at tree level in Eq.~\eqref{eq:hPT}  
provides a $\log (s)$ term in $w_N$:
\begin{align}
\frac{g^2 C_2}{2}\, \frac1{\pi^2} \log s\,.
\end{align}

For example, at $b=0.365$ and $N=11$, we obtained $0.3278$ for the
coefficient of the $\log S$ term in $w_N$ (cf.~Fig.~\ref{fig:c1SDepN11}). 
Matching the numerical result with PT would require $\frac{g^2 N}{4 \pi}\approx 
1.03$. This is consistent with the perimeter term determination of 
$\frac{g^2 N}{4 \pi}$.

\section{\boldmath $L\times 2L$ loops -- shape dependence}
\label{sec:L2L}

We turn now to a 
study of the shape dependence of the size-independent term 
in $w_N$ and compare it with the
effective-string prediction. In general, a shape-dependent parameter characterizing a loop $\C$ is
a dimensionless number describing $\C$ which is invariant under a 
scaling or an $\R^4$ space-time symmetry applied to $\C$. 
For rectangular $L_1\times L_2$ loops 
it is convenient to introduce the
modular invariant shape parameter
\begin{align}
\zeta=\frac{L_1}{L_2}+\frac{L_2}{L_1}\,.
\end{align}  
This $\zeta$ should not be confused with the $\zeta$-function that
will appear later.

The accuracy we now need does not permit taking the $N\to \infty$ limit.
We restrict our attention to the $N=7,11$ data. We shall see that the numbers we
compute are identical within errors for $N=7$ and $N=11$, indicating that 
it is unlikely that they will change in a substantial manner in the $N=\infty$ limit. 

At fixed $b$, $S$, $V$, and fixed
finite $N$, we expect
\begin{align}\label{eq:wL1L2}
w_N(L_1,L_2)+\frac 14 \log L_1 L_2 = c_{1,N}(\zeta) + c_{2,N}
\frac{L_1+L_2}{2} + \sigma_N L_1 L_2 + \mathcal
O\left(\frac 1{\sigma_N L_1 L_2}\right)\,.
\end{align}
Here, the arguments $b$, $S$, $V$ are omitted and the single length
scale $L$ of squares is replaced by $L_1$ and $L_2$ for rectangles. 

We extracted the lattice string tension from square $L\times L$ loops at fixed
finite $N$, $b$, $S$, $V$ in Sec.~\ref{sec:SigmaFinN}. 
We determined $\sigma_N$ by fitting the data using 
Eq.~\eqref{eq:SigmaFromSquaresFinN} 
with fit parameters $\sigma_N$ and $c_{2,N}$. After subtracting area and
perimeter terms, we fitted $w_N(L,L)+\frac 14 \log L^2-\sigma_N
L^2 - c_{2,N} L$ to a constant. This constant is now denoted by $c_{1,N}(\zeta=2)$.

We now analyze the results obtained for 
a sequence of rectangular loops at the same $b$, $S$, $V$,
$N$ with $L_2=2 L_1$, i.e., $\zeta=\frac 52$ fixed. 
Using the results for $\sigma_N$ and $c_{2,N}$ obtained from square loops 
above, we
determine $c_{1,N}(\zeta=\frac 52)$ by fitting
$w_N(L,2L)+\frac 14 \log\left( 2L^2\right)-\sigma_N
2L^2 - c_{2,N} \frac 32 L$ to a constant.
Figures~\ref{fig:N7-DeltaC1} and \ref{fig:N11-DeltaC1} 
show plots of $c_{1,N}(2.5)-c_{1,N}(2)$ as a
function of $b$ for $N=7$ and $N=11$. The $L$-ranges 
used for fitting $c_{1,N}$ are $6\leq L\leq 10$ for square
loops and $4\leq L \leq 7$ for $L\times 2L$ loops. The smallest
loop areas included in each set are therefore 36 and 32, respectively, putting them
close to the large-$N$ phase transition point.

\FIGURE[htb]{
\includegraphics[width=0.9\textwidth]{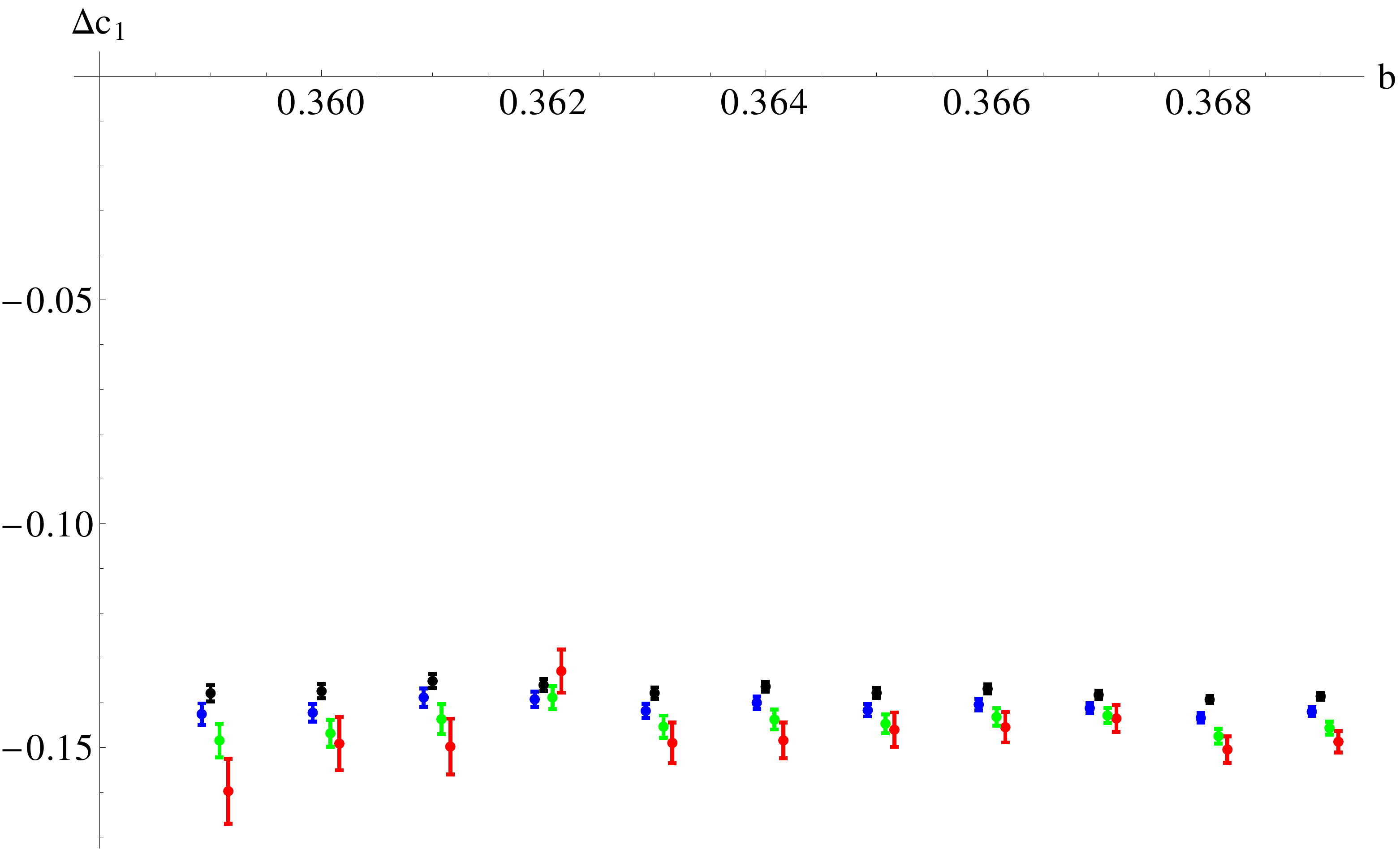}
\caption{Plot of $c_{1,N}(2.5)-c_{1,N}(2)$ for $N=7$ (on $V=24^4$) as a
  function of $b$ for $S=0.2$ (red), $S=0.28$ (green) $S=0.4$ (blue), and $S=0.52$ (black).} 
\label{fig:N7-DeltaC1} 
}

\FIGURE[htb]{
\includegraphics[width=0.9\textwidth]{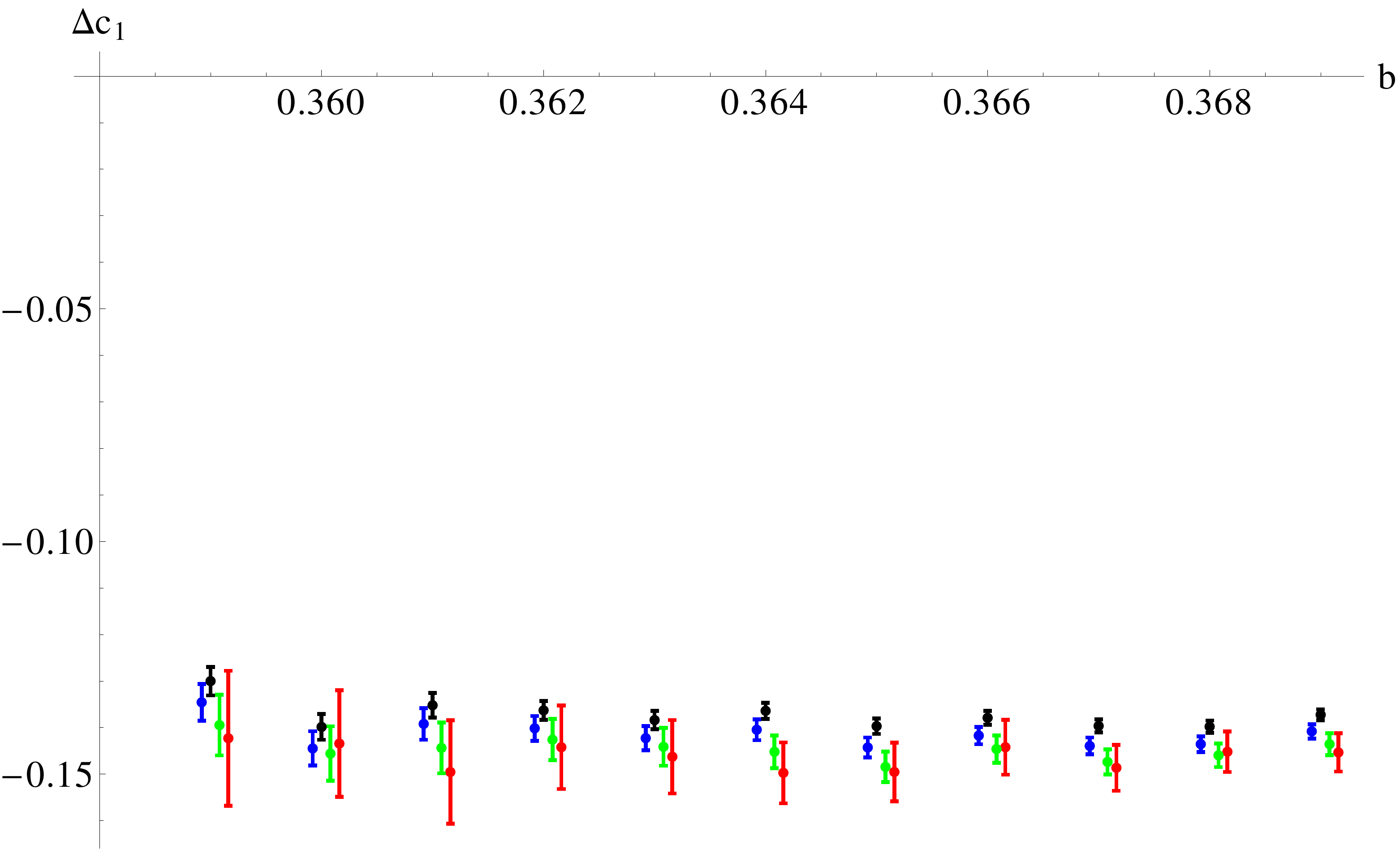}
\caption{Plot of $c_{1,N}(2.5)-c_{1,N}(2)$ for $N=11$ (on $V=18^4$) as a
  function of $b$ for $S=0.2$ (red), $S=0.28$ (green) $S=0.4$ (blue), and $S=0.52$ (black).} 
\label{fig:N11-DeltaC1} 
}

The effective-string prediction for $c_{1}(2.5)-c_{1}(2)$ is
\begin{align}
\frac 12 \log\left(\frac{\eta(2 i)\eta(i/2)}{\eta(i)^2}\right)\approx-0.08664\,, 
\end{align}
where $\eta(x)$ is the eta-function. We find that the effective-string prediction 
is smaller than the observed 
values (cf.~Figs.~\ref{fig:N7-DeltaC1}, \ref{fig:N11-DeltaC1}) by a
factor of about 1.5 to 1.7. Within statistical errors, our results for
$c_{1,N}(2.5)-c_{1,N}(2)$ do not depend on $b$, $S$, or $N$.

\section{Almost square loops}

We use sequences of almost square loops with sides $L_1=L$, $L_2=L\pm 1$ to 
cross check
our results for the string tension and the shape dependence of $c_{1,N}$. 
For these loops, the shape-parameter $\zeta$ changes with $L$ and is given by
\begin{align}
\zeta=\frac{L\pm 1}{L}+\frac{L}{L\pm 1}=2+
\frac1{L^2}\mp \frac1{L^3}+\mathcal O\left(L^{-4}\right)\,.
\end{align}

\subsection{String tension}

Expanding $c_{1,N}(\zeta)$ around $\zeta=2$, 
we obtain from Eq.~\eqref{eq:SigmaFromSquaresFinN} 
\begin{align}\label{eq:diffwRCT}
\frac 12 \left( w_N(L,L+1)-w_N(L,L-1)+\frac 14 \log
\frac{L+1}{L-1}\right)=\sigma_N L+\frac{c_{2,N}}2 +\ldots
\end{align}
We dropped corrections of order $\frac1{L^{3}}$ from the $\zeta$ expansion and
$\frac1{\sigma L^3}$ from the effective-string expansion. 

Similarly to the procedure for square loops, we first take $\lim_{N,V\to\infty}
w_N$ and then determine the infinite-$N$ string tension from $w_\infty$. 
Here, we use only method
1), i.e., we compute the limit $w_\infty$ from data at $N=7$, $11$, $19$, $29$ on
volumes $V=24^4$, $18^4$, $14^4$, $12^4$, 
respectively (for $N=7$ we use $V=24^4$ for
$b\geq 0.365$ and $V=20^4$ for $b\leq 0.364$). 

To determine $\sigma$ and $c_2$ from Eq.~\eqref{eq:diffwRCT} (at infinite $N$)
we use loops of sizes $6\times 7$, $7\times 8$, $8\times 9$.

The results for the infinite-$N$ string tension $\sigma(b)$ determined in this
manner agree very well
with those obtained from square loops, cf.~Table~\ref{tab:sigmasRCT} 
and Fig.~\ref{fig:Sigma-bRCT}.

\FIGURE[htb]{
\includegraphics[width=0.9\textwidth]{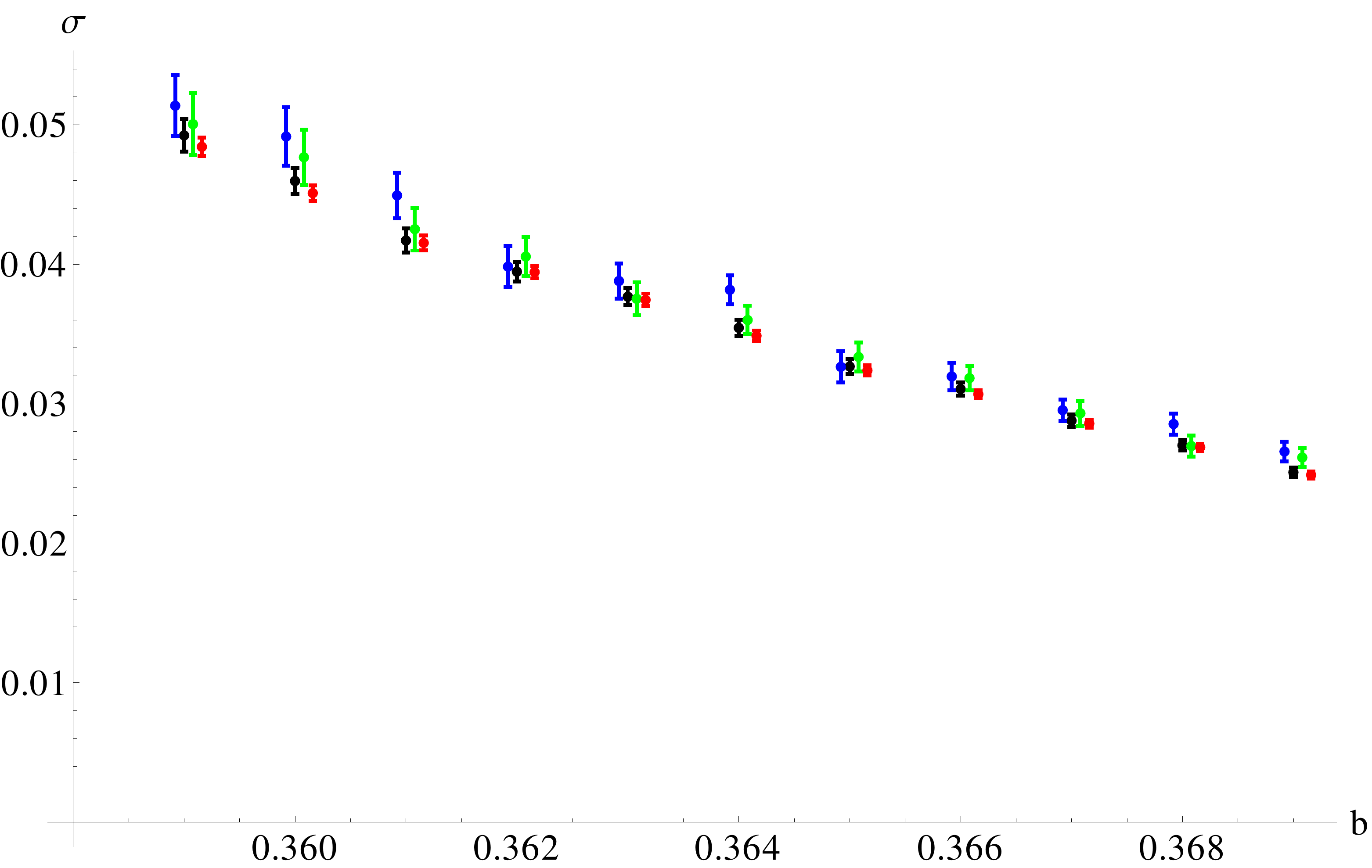}
\caption{Plots of the infinite-$N$ string
  tension obtained from rectangular $L\times L\pm1$ loops
  (cf.~Table~\ref{tab:sigmasRCT}) in red, together with results from square
  loops in black, blue, green (identical to plots in Fig.~\ref{fig:Sigma-b}).} 
\label{fig:Sigma-bRCT} 
}

\subsection{Shape dependence}
\label{sec:AlmostSqShape}

Using square and almost square loops, we can study the shape dependence of
$c_{1,N}(\zeta)$.
From Eq.~\eqref{eq:wL1L2}, ignoring corrections of order $1/L^4$ and $1/\sigma L^4$, 
we obtain 
\begin{align}\label{eq:c1prime}
\Delta w_N(L)\equiv w_N(L,L)-\frac12 w_N(L,L+1)-\frac 12
w_N(L,L-1)=-\frac{c_{1,N}'(2)}{L^2}-\frac1{8 L^2}+\ldots
\end{align}
In the first term on the r.h.s $c_{1,N}'(\zeta)\equiv\frac{dc_{1,N}(\zeta)}{d\zeta}$. 
The second term on the r.h.s.~results from the $\frac 14 \log L_1 L_2$
term in Eq.~\eqref{eq:wL1L2}. 
To determine $c_{1,N}'(2)$ we multiply $\Delta w_N$ by
$L^2$ and fit to a constant in the range $5\leq L \leq 8$ (for an example see Fig.~\ref{fig:C1pEx}). 
Note the constancy of $c_{1,N}'(2)$ as a function the gauge coupling. This indicates that
the number we extracted from the data is already in the 
continuum limit within quite small errors. 

In $\Delta w_N(L)$, both the shape-dependent constant $c_1(\zeta)$ and the
$\frac 14 \log L_1 L_2$ term result in terms of the order $1/L^2$ and only
their sum can be determined.  
From the effective-string prediction, we would expect $L^2 \Delta w_N(L) \to
0.0372764$ for large $L$  taking into account 
both contributions. This produces the value  $c_1'(2)\approx -0.162276$. 

The effective string
model produces asymptotic predictions for both  $c_1(\zeta)$ and $\frac 14 \log L_1 L_2$ 
from the determinant of  Gaussian fluctuations.  
One may refer to this prediction as a 1-loop
prediction of effective string theory, to distinguish 
it from subleading contributions, suppressed by
powers of the area in units of the string tension. 
 
Our numerical results for $L^2 \Delta w_N(L)$ deviate 
significantly from the asymptotic string
prediction (cf.~Fig.~\ref{fig:C1pEx}). There is no upfront 
indication in the data that subleading terms 
in the asymptotic series given by effective string theory play any role since there is no 
dependence on the gauge coupling. Such a dependence
would have to appear if the data were 
better described, say, by including a subleading correction. 
This subleading correction would
go as one over the area in string tension units and would 
depend on the gauge coupling $b$ 
through the lattice string tension $\sigma$. 
Here, we cannot decide whether the 
deviation originates from the shape-dependent constant $c_1(\zeta)$ or the
$\log L_1 L_2$ term. However, when we determine the string tension from square
loops, our results seem to be consistent with a $\log L$ term as predicted by the
string model (see also Sec.~\ref{sec:validationLog}). 
This indicates that $c_1(\zeta)$ is responsible for the
deviations.   
 
Taking into account the $\frac 1 {8 L^2}$ term coming from the $\log$ (i.e.,
we assume that the coefficient of the $\log L_1 L_2$ term in $w_N$ is correctly
determined by the string model), our results
for $c_1'$  exceed the asymptotic string prediction by a factor of about 1.6 to 1.8. 
Since $c_1(2.5)$ in the string prediction is very well approximated (to an
accuracy of 2\%) by an expansion around $\zeta=2$ to linear order, this
discrepancy is consistent with the discrepancy of $c_1(2.5)-c_1(2)$ observed
in Sec.~\ref{sec:L2L}. Our result here is also in agreement within errors with~\cite{tony} 
who independently report a deviation from string theory. 

We do not observe any significant dependence on $b$, $S$ or $N$
(cf.~Fig.~\ref{fig:C1p} for results at $S=0.4$).

\FIGURE[htb]{
\includegraphics[width=0.8\textwidth]{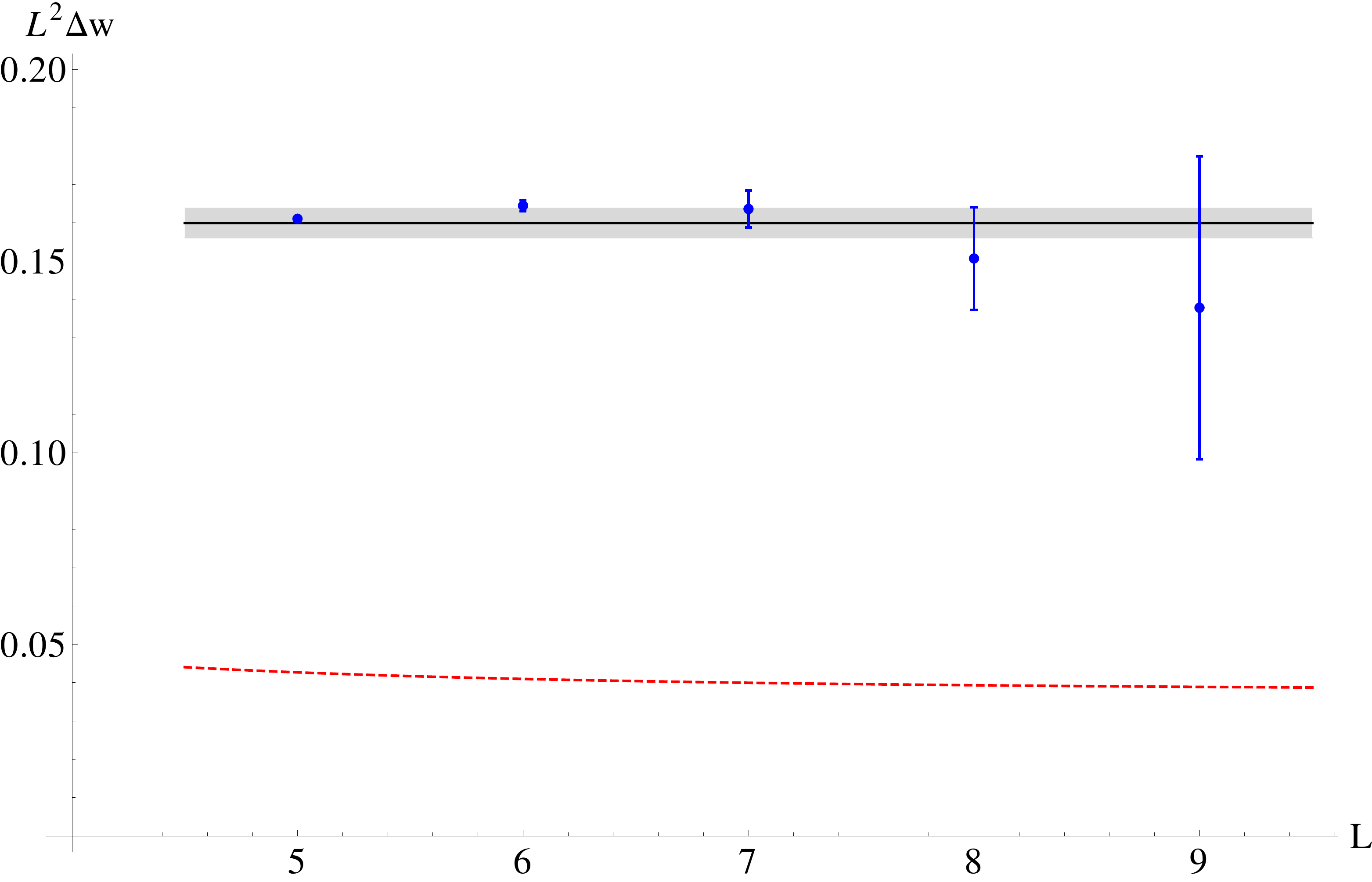}
\caption{Plot of $L^2 \Delta w_N(L)$ for $N=11$, $V=18^4$,
  $b=0.365$, $S=0.4$ (blue points), result of fit to a constant (black solid
  line, error estimate indicated by the gray band), and string prediction (red
dashed line).} 
\label{fig:C1pEx} 
}

\FIGURE[htb]{
\includegraphics[width=0.8\textwidth]{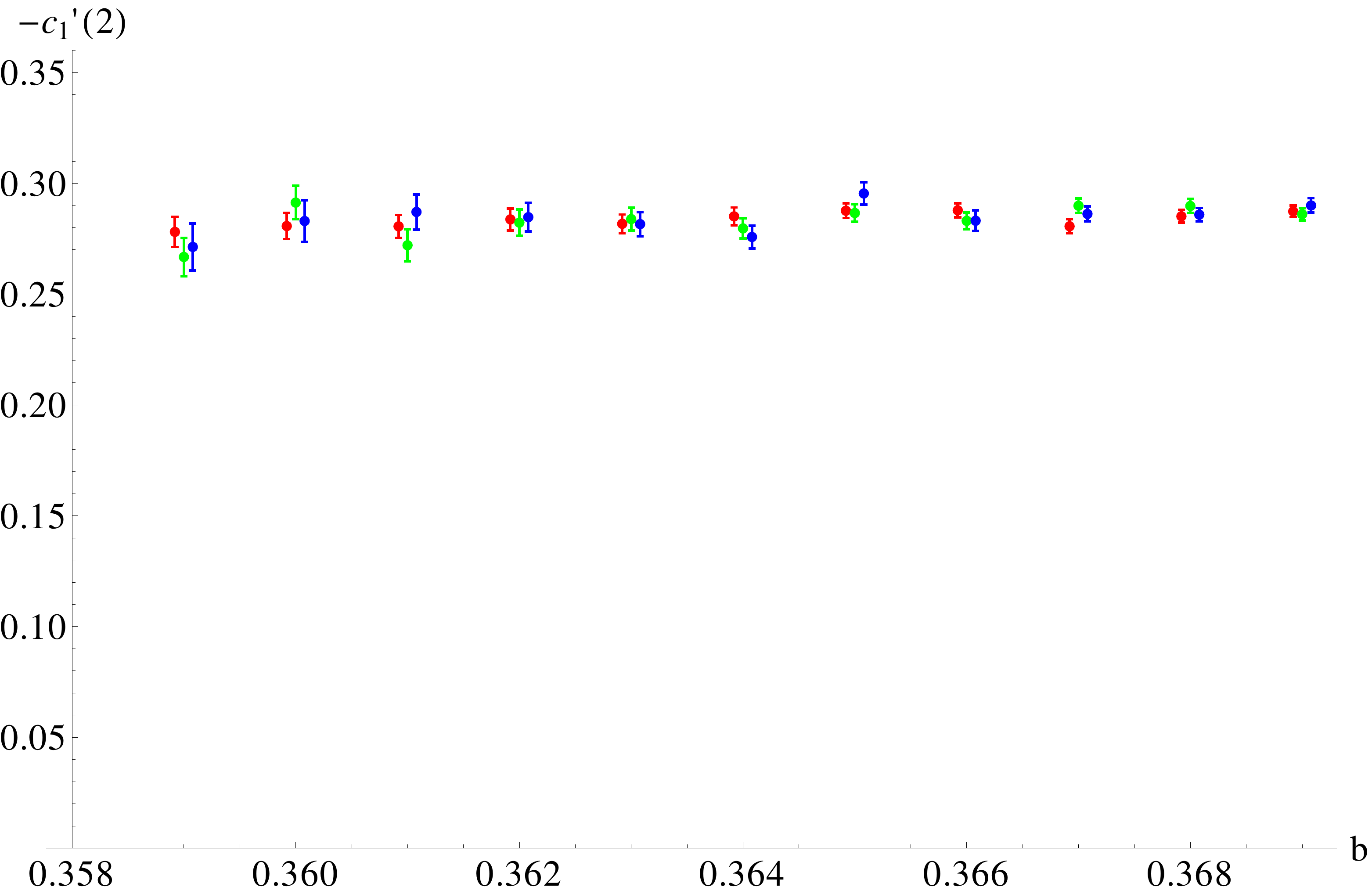}
\caption{Plot of $-c_{1,N}'(2)$ as a function of $b$ at $S=0.4$ for $N=7$ on
  $V=20^4$ (red), $N=11$ on $V=18^4$ (green), and $N=19$ on $V=14^4$
  (blue). (We define $c_1'(2)$ as $c_1'(2)=-L^2 \Delta w_N-\frac 18$; 
  the effective-string prediction is $c_1'(2)\approx -0.162276$.)} 
\label{fig:C1p} 
}

There is a question we shall address later on: Is it consistent within effective string theory 
to keep only the leading term? This question is meaningful in the sense that the theory 
predicts a correction of known, universal strength. The expansion in inverse area 
produces a series which likely is divergent. If a subleading correction is larger or equal
to the leading one, one would conclude that either further terms are needed or, one is
outside the reach of the asymptotic series in inverse size altogether and no further terms
would be of any help. Then the fact that the data showed no 
dependence on the gauge coupling $b$ 
would have to be explained in some other way, unrelated to 
effective string theory. To prepare
for this eventuality we need to address two questions first. 
Do we have numerical evidence for 
the logarithm of area term? What would perturbation theory 
have to say about the Wilson loops
in this range?

\subsubsection{Perturbation theory}

Extracting the shape-dependent terms from square $l\times l$ and almost square
$l\times(l\pm \delta l)$ loops similar to $\Delta w_N$ 
defined in Eq.~\eqref{eq:c1prime} in Sec.~\ref{sec:PT}, we obtain
\begin{align}
&\Delta w_N^{\text{PT}}(l,\delta l,s)=w_N^{\text{PT}}(l,l,s)-\frac 12
w_N^{\text{PT}}(l,l+\delta l,s)-\frac 12 w_N^{\text{PT}}(l,l-\delta l,s)\nonumber\\
&\quad=\frac{g^2 C_2}{2}\left(\frac 1{2\pi^2} \log\left(1-\frac{\delta l^2}{l^2}\right)+h_0(1)-\frac 12
h_0\left(1+\frac{\delta l}l\right)-\frac 12
h_0\left(1-\frac{\delta l}l\right)+\mathcal O\left(\frac s{l^2}\right)\right)
\nonumber\\
&\quad
=\frac{g^2 C_2}{2}
\left(
\left(-\frac12 h_0''(1)-\frac 1{2\pi^2}\right)\left(\frac{\delta
  l}{l}\right)^2+
\mathcal O\left(\frac s{l^2},\frac{\delta l^4}{l^4}\right)
\right)\,.
\end{align}
Numerically, we found $h_0''(1)\approx -0.260476$ and thus
\begin{align}
\Delta w_N^{\text{PT}}(l,\delta l,s)=\frac{g^2 C_2}{2}
\left(
0.0795774 \left(\frac{\delta
  l}{l}\right)^2+
\mathcal O\left(\frac s{l^2},\frac{\delta l^4}{l^4}\right)
\right)\,.
\end{align}

We have seen that some terms which diverge at zero smearing, 
and clearly are outside the reach of effective string theory, 
enter in the logarithm of the smeared Wilson loop in a simple additive manner. 
The case of shape dependence is more complicated and shall be discussed 
later on, when we look at possible explanations for the deviation of the shape
dependence we measure from the asymptotic prediction of effective string theory.

\section{All loops: validation of the log-term}
\label{sec:validationLog}

So far we have assumed that all our Wilson loops had
a prefactor given by $(\rm Area)^{\frac 14}$. 
We mentioned that attempts to carry out our fits without
this term produced substantially lower quality fits. 
We now would like to determine whether the power $\frac 14$
really is selected by our numbers. To do this we
need an amount of data and accuracy which does not allow 
us to consider separately different loop shapes or take
the infinite-$N$ limit. 
We do global fits to all our data at two values of $N$.

At fixed $N$, $b$, $S$, we fit square $L\times L$, 
almost square $L\times (L+1)$, and
rectangular $L\times (2L)$ loops to
\begin{align}
w_N(L_1,L_2)= \sigma L_1 L_2 + c_3 \log\left(L_1 L_2\right) +  c_2
\frac{L_1+L_2}{2}+c_1+c_4\left(\frac{L_1}{L_2}+\frac{L_2}{L_1}-2\right)\,. 
\end{align}
Some results are given in Tables \ref{tab:fullfitN7} and
\ref{tab:fullfitN11}. Going to smaller loops, $\chi^2/N_{\text{dof}}$ starts
to increase significantly. 
Tables \ref{tab:fullfitN7} and \ref{tab:fullfitN11} tell us that the
value $ \frac 14 $ for the exponent of the area is 
consistent with the data. 

\subsection[$\log(\text{Area})$ 
versus shape dependence]{\boldmath $\log(\text{Area})$ versus shape dependence}

We have seen now that one prediction coming from the 
determinant in the effective string description works close to the 
large-$N$ transition in the eigenvalues and the other does not.

These two predictions are somewhat 
different even within effective string theory. The determinant of
the small fluctuations of the spanning surface around the minimal
area one is most conveniently evaluated using $\zeta$-function 
regularization. The determinant itself is ill defined and
$\zeta$-function regularization provides one way to extract finite
universal features. Only such features are conceivably relevant to Wilson loops. 

Within $\zeta$-function regularization one has the option to make a decision
about how to treat the directions perpendicular to the surface which now
are fields in the two-dimensional theory on the world-sheet. This description
is supposed to be geometrical without introducing any scale. It is
convenient to enforce this by thinking in terms of two-dimensional 
gravity. Then, it is natural to view the fields as half-densities \cite{hawking-zeta}.
With this convention, the power of the area is a regulated number of
degrees of freedom, coming from $\zeta(0;D)$ where $D$ is the fluctuation
operator. The rest of the determinant is just a function of the
shape parameter $\zeta$ and
comes from the derivative $d\zeta(w;D)/dw|_{w=0}$, reflecting the eigenvalues
of $D$ more directly. 

$\zeta(0,D)$ has additive contributions coming from each kink in 
our planar curve $\C$:
\begin{align}
\sum_i\frac{\pi^2-\theta_i^2}{24\pi \theta_i}\,.
\end{align}
Here, the $0\le\theta_i\le2\pi$ are the angles at each 
kink measured by an arc contained
in the interior of $\C$. For a rectangular loop, summing over the 
two orthogonal directions to the surface produces the factor $\frac 14$.
For a backtracking loop the corner term would blow up. For
a smeared Wilson loop, a backtracking segment makes no contribution.

$\zeta$-function regularization extracts the 
universal predictions. It is natural to use 
a two-dimensional lattice regularization for the effective string instead. 
This is so because in the strong coupling expansion of the lattice gauge 
theory one can identify contributions given by the exponent of the area
of a spanning surface made out of tiles that can be labeled by two fields
depending on two coordinates on a square world-sheet lattice. 
It is easy to numerically determine the asymptotic expansion in $L$
of the fluctuation determinant for a square loop,  
assuming the two fields to be continuous: 
\begin{align}
\log\prod_{n=1}^{L-1}\prod_{m=1}^{L-1} 
\left ( 2-\cos\frac{n\pi}{L}-\cos\frac{m\pi}{L}\right )
\sim 0.4731 L^2 -0.37645 L -\frac 12 \log L -0.09039+\ldots
\end{align}
We see that there is an area term, a perimeter term and 
a constant but they
are absorbed into the physical area law, the well defined perimeter
 and constant terms in the case of smeared loops. 
 It is just as easy to do this for a rectangular loop.
 It is possible to derive the asymptotic expansion for rectangular loops by 
 analytical means too. Below we reproduce part of 
 Eq.~(4.20) from \cite{david}:
 \begin{align}
 \log\prod_{n=1}^{N-1}\prod_{m=1}^{M-1} 
\left ( 4-2\cos\frac{n\pi}{L}-2\cos\frac{m\pi}{L}\right )
&\sim MN\frac{4G}{\pi}-(M+N) \log(1+\sqrt{2})\nonumber\\
&\quad -\frac{1}{4} \log MN +...
 \end{align}
 Here, $G=1-1/3^2+1/5^2...$ is Catalan's constant. The universal results 
 using $\zeta$-function regularization are reproduced. The derivation of~\cite{david}
 shows that the logarithmic term comes from modes that vary little (have small
 wave numbers) in one of the two directions parallel to the sides of the rectangle. 
 These modes are the ones most affected by the Dirichlet boundary conditions. 
 
 Looking at the exact expression for the determinant for finite integer $(M,N)$ 
 we find that the asymptotic expansion truncated after the constant term provides
 estimates for the logarithm of the determinant at relative accuracy 0.014\% for
 $M,N\ge 5$ and 0.0065\% for $M,N\ge 6$. The largest deviation in both sets is
 at $M=N$. For loops at fixed aspect ratio of 2, $M=2N$, the relative
 accuracy is $0.024\%$ for $N=4$ and $0.0012\%$ for $N=8$. In this simple case, 
 quite small loops are very well described by the asymptotic series 
 without any terms that vanish as $M,N\to\infty$. 
 
 One cannot tell a priori what if
 anything is left from this numerical observation when one looks at real Wilson loops. 
 If this held also for real Wilson loops we would expect to see a shape dependence
 independent on the gauge coupling $b$ for loops as small as we looked at. But, then,
 the number should have agreed with the asymptotic prediction of the determinant. 
 
 \subsection{Possible explanations of the deviation of the shape dependent 
 constant from the prediction of asymptotic effective string theory}
 
One employs effective string theory under the hypothesis that once the 
ultimate asymptotic regime is
entered, it will take complete control of the shape
dependence of the functionals $W(\C)$. Specifically, 
in the kind of asymptotic expansion in
dilatation that we are considering (which is different 
from looking at the separation dependence
of the interquark force for example) one has,
as $\rho\to\infty$, the following behavior for a dilated loop $\rho \C$:
\begin{align}\label{eq:asympt_exp}
 \log(W(\rho\C)) &\sim -\sigma\rho^2 {\rm Area}_{\rm min}(\C)+ 
 \Gamma_P \;\rho\,{\rm Length}(\C) + \Gamma_1(\C)\log(\rho) +\Gamma_2(\C)
 \nonumber\\&\quad + \Gamma_K (\C)+ \Gamma_3(\C)/\rho^2 +
\Gamma_4(\C)/\rho^3+\Gamma_5(\C)/\rho^4+{\cal O}(1/\rho^5)\,.
\end{align}

One can think about $\rho$ and $\C$ as follows: Let the minimal area with some
$\C'$ as boundary be unique. Multiply this area by the string tension and get 
a pure number. Scale $\C'$ by an amount that makes this number equal to 1 
and call the so obtained curve $\C$. Choose a parametrization of this curve
$\C$ by a $x_\mu(\tau)$ with a choice of $\tau$ such that $\dot x^2=1$. 
Then, the information contained in $\C$ is equivalent to the information 
contained in the set of all global $\O(4)$ invariants one can
construct out of the function $\dot x_\mu$. $\tau$ goes once round $\C$.
Note that with this convention, the perimeter of $\rho\C$ depends on $\C$.

$\C$ is allowed to have kinks. $\sigma$ is dimensional and has nothing to
do with effective string theory, except that $\sigma >0$. The perimeter coefficient $\Gamma_P$ is a non-universal number
independent of $\C$. 
$\Gamma_1(\C)$, $\Gamma_2(\C)$, $\Gamma_3(\C)$, and $\Gamma_5(\C)$ 
are scale-invariant functions of $\C$
and universal. 
$\Gamma_4(\C)$ is a scale-invariant function of $\C$ with 
one non-universal 
overall constant multiplicative factor. 

\begin{align}
\Gamma_K(\C)=\sum_{\rm kinks} F (\dot x_+ \cdot \dot x_- |_{\rm kink})\,.
\end{align}
A crucial assumption is that the non-universal function $F()$ is independent of $\C$, 
and that its argument is given by the discontinuity in the tangent of $\C$ at the kinks. 
Without this assumption one cannot separate $\Gamma_2(\C)$ from $\Gamma_K(\C)$.
With this assumption the $\Gamma_K$ term can be eliminated by comparing loops $\C_1$
and $\C_2$ which have the same set of kinks. In this case, effective string theory  
makes a testable prediction for 
$\Gamma_2(\C_1 )-\Gamma_2 (\C_2 )$. 

The presence of the terms $\Gamma_P$ and $\Gamma_K$ in Eq.~\eqref{eq:asympt_exp} can 
be motivated in four-dimensional Yang-Mills field theory. The Wilson loop has perimeter
and corner ultraviolet divergences whose removal will introduce some ad-hoc parameters 
one could not expect effective string theory to know about. According to this logic, Wilson 
loops in three-dimensional Yang-Mills theory with continuous gauge groups would not
require a $\Gamma_K$ term. From the effective string theory point of view there is no 
motivation for such distinction between four and three dimensions. 

Equation~\eqref{eq:asympt_exp} is tested in only a limited manner on rectangular loops.
$\Gamma_K$ is a fixed constant in this set of loops.
Since we encountered a deviation at order $\rho^0$ at the 
level of the universality of $\Gamma_2(\C)$, 
we could suspect that the problem has to do with the presence of kinks. 
Since any kink can be rounded, 
the requirement on the size of $\rho$ needed in order 
to justify keeping only the leading terms up to and including $\Gamma_2$ 
in the asymptotic expansion in large $\rho$ 
involves the local radius of curvature of $\C$, ${\cal R}$.
For a kink-free $\C$, parametrically described by a $x(\tau)$, we have
\begin{align}
{\cal R}(\tau)=\frac{({\dot x}^2)^{\frac 32}}{\sqrt{{\dot x}^2{\ddot x}^2-
({\dot x} \cdot {\ddot x} )^2}}\,.
\end{align}
${\cal R}$ is invariant under re-parameterizations of the boundary and under
$\C\to\rho\C$ it scales like $\rho$. 
With the choice ${\dot x}^2=1$, we have 
${\cal R}=1/|{\ddot x}|$, which endows the parameter $\tau$ with the same
dimension as that of $x$. 
Effectively, in 
the presence of a single ``almost''  kink, the Wilson loop depends on two hugely 
disparate scales. There exists a field-theoretical definition of the 
smeared Wilson loop, and we know from experience that observables depending on 
very disparate scales are hard to calculate. It could be though that the effective string theory
provides a framework which is so different from field theory 
that this experience is irrelevant.

We now proceed to ask a simpler question: does effective string theory tell us that the 
subleading corrections to the shape dependence are so large that we had no right
to compare the data to the leading asymptotic prediction? Had we obtained agreement, 
we probably would not have raised this question, like in many previous studies of the 
interquark force~\cite{necco_sommer}, which produced 
agreement in the leading term in the 
asymptotic expansion relevant to that case.

Within the premise of effective string theory we are working, there is no
substantial difference for rectangular loops 
between something as simple as $Z_2$ gauge theory in 
three Euclidean dimensions and $\SU(\infty)$ pure gauge theory in four Euclidean 
dimensions. 
Consequentially one has ready 
examples in the recent literature~\cite{turin} for how to include
subleading terms. 

In the next subsection we carry out this exercise on our data.

After that we come back to discuss at a more intuitive level possible differences 
between $Z_2$ three-dimensional gauge theory on the lattice, which is exactly dual to
the three-dimensional Ising model and hence has a field-theoretical continuum description
built around the Wilson-Fisher fixed point, and four-dimensional planar $\SU(N)$ gauge theory.

\subsection{Subleading terms in effective string theory}
\label{sec:subleading-EFT}

The subleading corrections that have been looked at in detail come in at orders 
$\frac{1}{\rho^2},\frac{1}{\rho^3},\frac{1}{\rho^4}$ corresponding, respectively, to a bulk, a boundary and another bulk correction. Corner corrections have 
not been discussed in the effective string literature, as far as we know.
The two bulk corrections have universal coefficients, known functions of the shape parameter $\zeta$. 
The boundary term
has an adjustable coefficient. There exists an unresolved discrepancy in the 
coefficient of $\rho^{-2}$: 
there are two candidates, denoted as ${\cal L}_2 (u)$ and $\hat{\cal L}_2 (u)$ 
differing by a 
$u$-independent number, where $u+u^{-1}=\zeta$.\footnote
{We use ${\cal
  L}_2(u)=\left(\frac\pi{24}\right)^2\left(4u^2E_4(iu)+2E_2(iu)E_2(i/u)\right)$
and $\hat{\cal L}_2 (u)={\cal L}_2(u)-\frac{3}{32}$ as defined in
\cite{turin}.}
 In the plots we show the contributions of each candidate, hoping that one is correct.\footnote
{We assume a typographical error in the Fourier 
expansion of the Eisenstein series in
  \cite{turin} (Eq.~(A.4) in JHEP {\bf 1201}, 104)  and
instead use the expression 
$E_{2k}(i u)=1+\frac{2}{\zeta(1-2k)}\sum_{n=1}^\infty \sigma_{2k-1}(n) \, e^{-2 \pi n u}$. Here, $\sigma_m(n)$ denotes the sum of
the $m$-th powers of the divisors of $n$. This corresponds 
to a change by a factor of
2 in the definition of $u$. 
}
 We have not
rechecked the calculations of the coefficients by ourselves.
We could 
add these terms as corrections to the effective string theory partition function
and then take the logarithm or directly to the logarithm. Since the 
numbers differ for our values of loop area, we show both cases.


We show different forms of contributions up to order $\rho^{-2}$ 
in three examples in 
figures~\ref{fig:NEW-C1pEx},~\ref{fig:NEW-Lx2L-Ex}, and~\ref{fig:NEW-Lx2L-Ex-Vs-A}. 
The black points/lines represent the numerical data in all figures. With
\begin{align}
\tilde w(L_1,L_2)\equiv w(L_1,L_2)+\frac 14 \log L_1L_2-\sigma L_1 L_2-c_2\frac{L_1+L_2}2 - c_1(\zeta=2)
\end{align}
for rectangular $L_1 \times L_2$ loops, the colored lines 
pertaining to the effective string description are defined as follows:
\begin{itemize}
\item Red: $\rho^0$-term coming from the determinant, $\tilde w(L_1,L_2)= \frac 12 \log\left(\eta(iu)\eta(i/u)\eta^{-2}(i)\right)$.
\item Solid green:  $\tilde w(L_1,L_2)= \frac 12 \log\left(\eta(iu)\eta(i/u)\eta^{-2}(i)\right) - \log\left(1+\frac{{\cal L}_2(u)}{\sigma L_1 L_2}\right)$ is used to include the term of order $\rho^{-2}$.
\item Dashed green: $\tilde w(L_1,L_2)= \frac 12 \log\left(\eta(iu)\eta(i/u)\eta^{-2}(i)\right) - \log\left(1+\frac{\hat{\cal L}_2(u)}{\sigma L_1 L_2}\right)$ is used to include the term of order $\rho^{-2}$.
\item Solid orange: $\tilde w(L_1,L_2) = \frac 12 \log\left(\eta(iu)\eta(i/u)\eta^{-2}(i)\right) - \frac{{\cal L}_2(u)}{\sigma L_1 L_2}$ is used to include the term of order $\rho^{-2}$.
\item Dashed orange:  $\tilde w(L_1,L_2)= \frac 12 \log\left(\eta(iu)\eta(i/u)\eta^{-2}(i)\right) - \frac{\hat{\cal L}_2(u)}{\sigma L_1 L_2}$ is used to include the term of order $\rho^{-2}$.
\end{itemize}
For the subleading terms, we use $\sigma=0.02916$ (as obtained from square loops for $N=11$ at $b=0.365$, cf.~Table~\ref{tab:sigmaN}).
Note that
\begin{align}
\Delta \tilde w(L)\equiv \tilde w(L,L)-\frac 12 \tilde w(L,L+1)-\frac 12 \tilde w(L,L-1)=\Delta w(L)-\frac 18 \log\left(1-L^{-2}\right)\,.
\end{align}
The adjustable order $\rho^{-3}$ term is not shown in the figures because we would have to fit
its coefficient. We also ignore the order $\rho^{-4}$ term, although it would produce
distinguishable numbers on the plots\footnote{We did not manage to reproduce the plots in~\cite{turin} which include the $\rho^{-4}$ term (but apparently ignore
the $\rho^{-3}$ term) when we simply 
implemented the equations therein. We did not pursue this issue 
because we felt that adding more clutter into our figures would be more harmful than 
informative.}

In $\tilde w(L,2L)-\tilde w(L,L)$ (cf.~Fig.~\ref{fig:NEW-Lx2L-Ex}), square loops 
are subtracted from rectangular loops which have an area twice as large. 
Therefore, the effective string predictions with ${\cal L}_2$ and $\hat{\cal L}_2$ 
differ at next-to-leading order. 

It is quite clear that if we wanted to add the missing $\mathcal{O}\left(\rho^{-3}\right)$ boundary term,
we could get agreement between theory and a few more data points, for smaller loops. 
One may conclude then that effective string theory applies to our data. That
the boundary term would scale correctly is an automatic consequence of 
the $b$-independence of the data. 

Alternatively, one may simply conclude that for our loops
effective string theory makes no definite prediction for the shape dependence.
We face a standard situation with asymptotic series: adding too many terms is
a bad idea and so is having too few. How one looks at the data becomes quite
subjective. 

Be that as it may, 
since the main focus of our work is to determine 
which predictions of effective string theory hold close enough to the 
infinite-$N$ phase transition and therefore could come in when one tried to match onto
perturbation theory,  we are left where we were before carrying out this exercise: 
beyond the area term we can use with some confidence also the logarithmic term. 
Employing effective string prescribed terms of higher order in $\rho^{-1}$ becomes 
questionable. 

\FIGURE[htb]{
\includegraphics[width=0.75\textwidth]{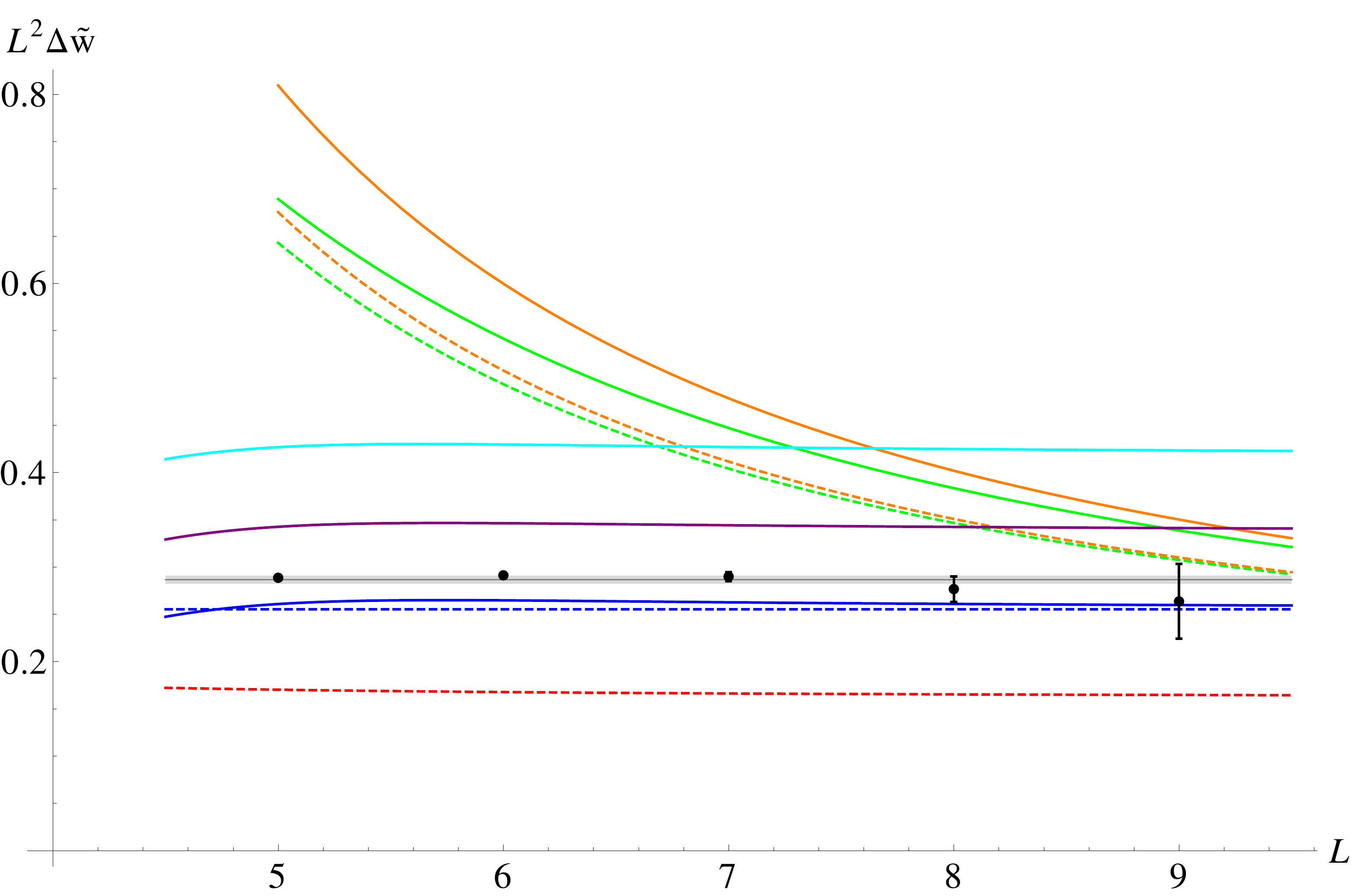}
\caption{Plot of 
$L^2 \Delta \tilde w(L)$ (black points) for $N=11$, $V=18^4$, $b=0.365$, $S=0.4$ together with the result of a fit to a constant (thin dark-grey
  line, error estimate indicated by the light-gray band). The various colored lines show estimates obtained from effective string theory (green, orange) and tree-level perturbation theory (blue, purple, cyan) as described in Secs.~\ref{sec:subleading-EFT} and \ref{sec:estimatesPT}.} 
\label{fig:NEW-C1pEx} 
}

 \FIGURE[t]{
 \includegraphics[width=0.75\textwidth]{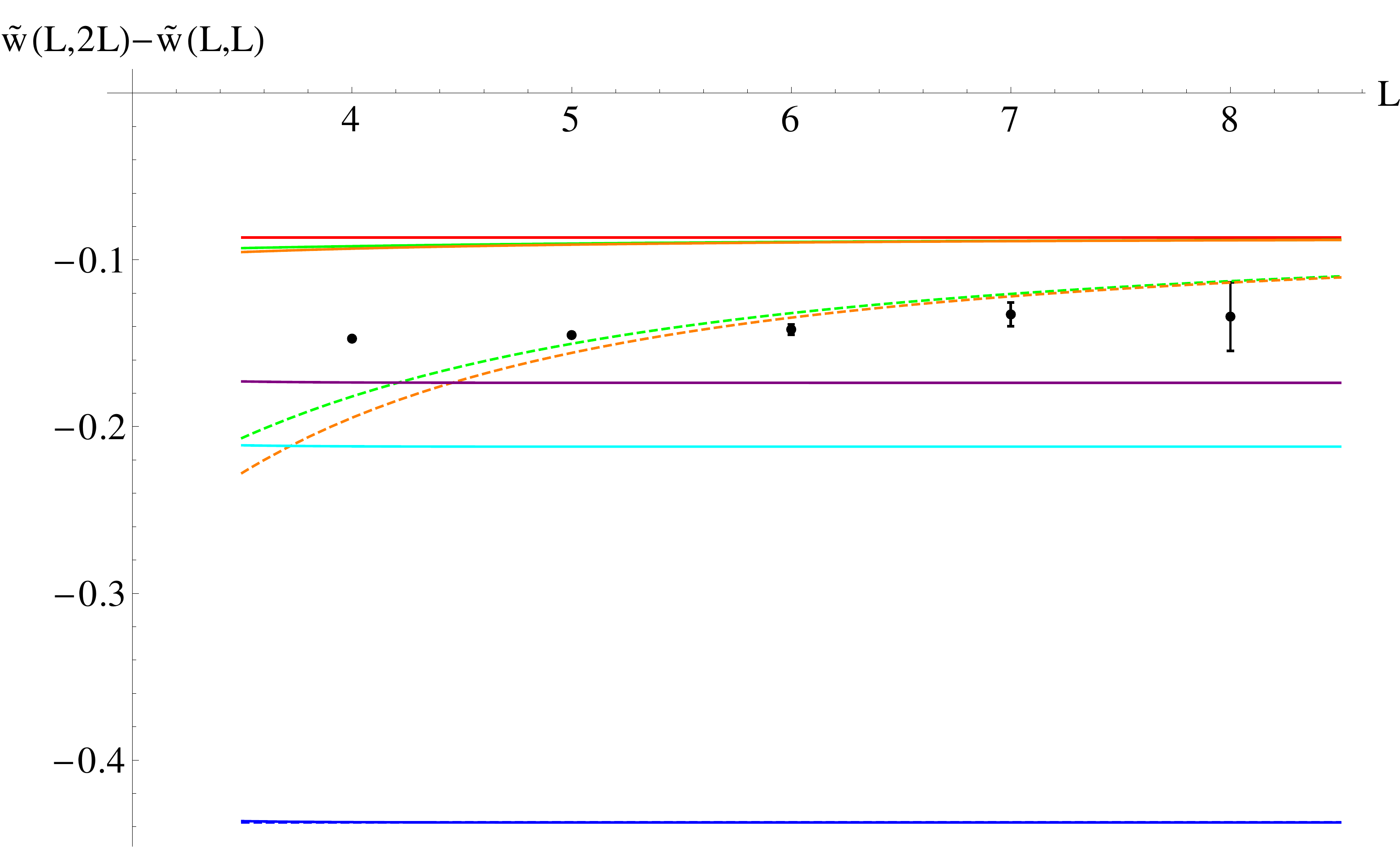}
 \caption{Plot of $\tilde w(L,2L)-\tilde w(L,L)$ (black points) for $N=11$, $V=18^4$, $b=0.365$, $S=0.4$. Numerical results for $\tilde w$ are obtained by 
 subtracting area-, perimeter-, and constant-term (with coefficients 
 determined from square loop data) from measured $w(L_1,L_2)+\frac14\log(L_1 L_2)$.} 
 \label{fig:NEW-Lx2L-Ex} 
 }

 \FIGURE[b]{
 \includegraphics[width=0.75\textwidth]{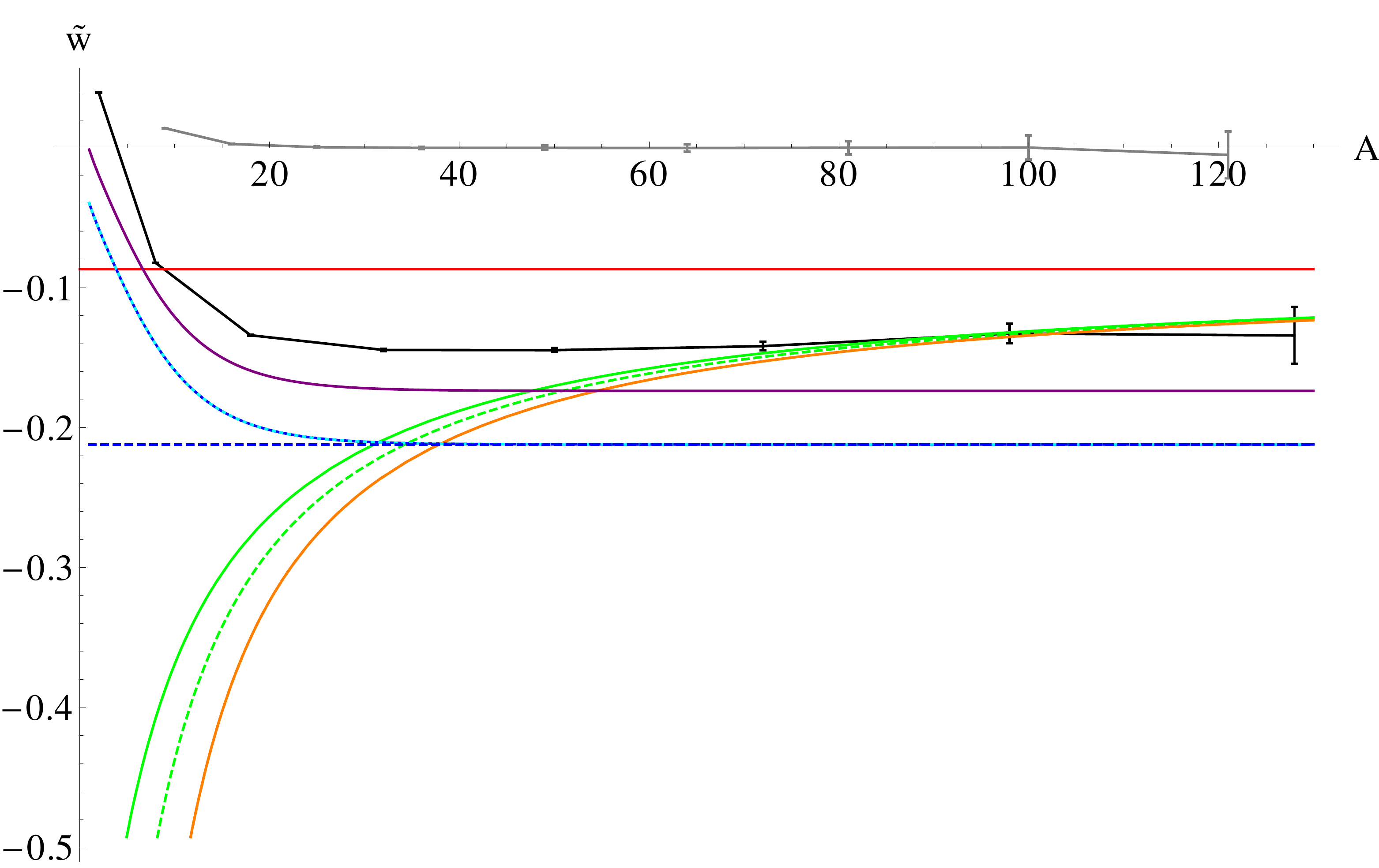}
 \caption{Plot of $\tilde w(L,2L)$ (black) and $\tilde w(L,L)$ (gray) 
 as a function of the area $A$ for $N=11$, $V=18^4$, $b=0.365$, $S=0.4$. Since $\tilde w(L,L)$ is zero within errors 
 (except for the smallest loops), $\tilde w(L,2L)$ can be viewed as a 
 reasonable estimate for $\tilde w(\sqrt{2A},\sqrt{A/2})-\tilde w(\sqrt{A},\sqrt{A})$. 
 The various colored lines show effective-string and perturbative predictions for this 
 difference as a function of the area $A$.} 
 \label{fig:NEW-Lx2L-Ex-Vs-A} 
 }

\subsection{Rough estimates in perturbative field theory}
\label{sec:estimatesPT}

We now take an orthogonal view: we try to see how the data could be described 
by field theory in perturbation theory. Just as with the effective string description, 
we are most likely outside the proper reach of this expansion too. It is
clear however that one cannot dispense with the area term, albeit that it is
not predictable by field-theoretical perturbation theory. Regarding the 
term logarithmic in the area coming from effective string theory, we choose
to eliminate it. Perturbation theory will come with its own logarithms and
there is no objective way to mix them with logarithms coming from effective string theory.
The relevant lines in the figures~\ref{fig:NEW-C1pEx},~\ref{fig:NEW-Lx2L-Ex},~\ref{fig:NEW-Lx2L-Ex-Vs-A} are defined below.
\begin{itemize}
\item Solid blue: $\tilde w(l_1,l_2,s) =w^{\text{PT}}(l_1,l_2,s)-
\frac{g^2 C_2}{2}\frac1{(2\pi)^{\frac 32}}\left(\frac{l_1+l_2}{\sqrt{s}}\right)$ 
(cf.~Eqs.~(\ref{eq:wNPT}, \ref{eq:hPT})). \newline
Note that shifting $\tilde w$ by a ($l_1/l_2$-independent) constant has no effect on the observables we are considering here. 
\item Dashed blue: leading term (in the large-$l_{i}/\sqrt{s}$ expansion) of the  above.
\item Cyan: $\tilde w(l_1,l_2,s)=w^{\text{PT}}(l_1,l_2,s)-\frac{g^2 C_2}{2}
\left(\frac1{(2\pi)^{\frac 32}}\left(\frac{l_1+l_2}{\sqrt{s}}\right)+
\frac1{\pi^2}\log\left(\frac s{l_1l_2}\right)\right)$. \newline
Replacing the $\log(s/l_1l_2)$-term in $w^{\text{PT}}(l_1,l_2,s)$ by $\log(s \Lambda^2)$, this term would no longer contribute to the shape dependence. Then the leading order contribution is determined by $h_0(l_2/l_1)$ only (cf. Eq. (4.24)).
\item Purple: $\tilde w_N=w_N^{\text{PT}}(l_1,l_2,s)-\frac{g^2 C_2}{2}
\left(\frac1{(2\pi)^{\frac 32}}\left(\frac{l_1+l_2}{\sqrt{s}}\right)+\frac1{\pi^2}
\log\left(\frac s{(l_1+l_2)^2}\right)\right)$.\newline
This corresponds to the replacement $\log(s/(l_1+l_2)^2) \to \log(s \Lambda^2)$ in $w^{\text{PT}}(l_1,l_2,s)$.
\end{itemize}
For our perturbative estimates shown in Figs.~\ref{fig:NEW-C1pEx},~\ref{fig:NEW-Lx2L-Ex},~\ref{fig:NEW-Lx2L-Ex-Vs-A}, we set the coupling constant to $\frac{g^2N}{4\pi}=1.03$, the value that we obtained from the coefficient of the $\log(s)$-term for $b=0.365$, $N=11$ (cf.~Sec.~\ref{sec:coeff-log-s}). 

Our information from perturbation theory is clearly too limited at this stage to draw any 
concrete conclusions. As far as we went, it seems to work just as well or as badly
as effective string theory.

\subsection{Suggestions for further research of the shape dependence}

Summarizing the situation so far, it seems that for moderate sizes loops one
observes a continuum shape dependence which might be explained by perturbation 
theory and might upon extension to much larger loops transit to another value given
by effective string theory. 

Taking into account what we know about nonabelian four-dimensional theories, we
think that the issue of shape dependence deserves further study, albeit somewhat
tangential to our own long-range project. 

Further numerical checks could be made focusing on the specific issue 
of shape dependence. An interesting set of Wilson loops amenable to 
study on hypercubic lattices have $\theta_i$ angles equal to $\frac{3\pi}2$.
The loop is not convex and for such angles the difference between the 
effective-string shape dependence and the field-theoretical one is enhanced.
Physically, in perturbation theory gluons exchanged between different segments
of the loop will predominantly choose to travel through the ``outside'' for an obtuse corner angle.
On the other hand, surfaces of effective string theory fluctuate above the ``inside''
of the loop, convex or not. In this context it would be also of interest to
consider self-intersecting loops given by fitting a figure of 8 onto a hypercubic lattice. 

We suggest that the shape dependence of planar Wilson loops presents 
an interesting case for testing the limitations of the effective string approach.
We know that a high-energy scattering event in QCD produces after
degradation into the IR a pattern of jets that is an imprint of perturbation theory.
\begin{itemize}
\item Could it be that even an asymptotically dilated 
Wilson loop with kinks in Euclidean space 
would have elements of shape dependence that are determined by the
field theory at short distances and which do not get washed out
at large distances? 
\item Were that the case, is the effective string theory framework flexible enough to
allow the inclusion of specific kink terms that can be adjusted to exactly reproduce 
the angle dependence of the anomalous dimensions associated with kinks? 
\end{itemize}

\subsection[How much should one expect 3D lattice $Z_2$ 
gauge theory to teach us about $\SU(\infty)$ 4D pure Yang-Mills theory?]{\boldmath How much should one expect 3D lattice $Z_2$ 
gauge theory to teach us about $\SU(\infty)$ 4D pure Yang-Mills theory?}

In 3D gauge theories with continuous groups there are no corner singularities. 
The perimeter term is only logarithmically divergent. The renormalization 
properties of the Wilson loop operators are significantly different. We emphasized 
several features of the corner singularities in 4D gauge theory. They have no analogues
in 3D. The corner singularities play a role in determining the shape dependence of
Wilson loops in 4D. Further, the case of discrete gauge groups in 3D is substantially
distinct from the case of continuous groups.

One place one can compare the two theories would be on the lattice. For 4D Yang-Mills
theory one has well defined loop equations. These equations have a formal 
continuum limit in which loops with kinks play a crucial role. The Ising model, which
is dual in three dimensions to the $Z_2$ gauge model also has a lattice loop
equation in terms of a ``barbed wire'' loop boundary~\cite{polyakov_book}. 
The equation is very different
from the four-dimensional lattice loop equation for Yang-Mills. 

These equations look stringy, and it has been the folklore that they would lead to an
exact dual string description. It seems plausible that if such an exact dual 
description exists, the appropriate effective string would bear some closeness to it,
say in terms of the correct massless degrees of freedom one should use. This may
make no difference at leading order, but at higher order what is local on the world
sheet in one description might be non-local in another. The string dual of the Ising 
model seems not to have a tunable string coupling constant, while the string
dual of $\SU(N)$ gauge theory seems to have one, which can be set to zero by
taking $N$ to infinity. There seems to be no analogous freedom and 
limit in the 3D case. Even if handles are exponentially suppressed for the three-dimensional 
Ising case, there is no control on this and no way we know of to actually estimate their 
numerical values; once one works at loop sizes and 
accuracies sensitive to higher order terms in $\rho$ than the leading one 
it is difficult to asses how much of a match between data and theory one ought to expect. 
In the large-$N$ limit of $\SU(N)$ gauge theory there is a credible argument that at least
one does not have to worry about handle corrections. 

We urge caution in drawing conclusions from 3D $Z_2$ lattice gauge
theory about 4D Yang-Mills theory. The results of applying effective string theory
to the three-dimensional Ising model are quite impressive in themselves, without
necessarily being relevant in detail 
to four-dimensional gauge theories of the type we 
have in Nature at a rather fundamental level.

\section{Conclusions}

A stringy parametrization for rectangular 
Wilson loops holds relatively well all the way
down to the large-$N$ transition point. 
Notably, the scale dependence of the $\Gamma_1(\C)\log(\rho)$-term is consistent with 
a correct interpretation of the number of degrees of 
freedom living on the worldsheet in the Nambu-Goto case.
However, when one gets to the subleading 
level of the dependence on purely shape-dependent parameters, 
the situation is less clear.

We are specifically interested in
the large-$N$ limit of four-dimensional $\SU(N)$ pure gauge theories
and in contours with corner singularities. It could be that there is
another effective string theory prescription that takes into 
account the corners in a special way. It would describe the 
large-$\rho$ asymptotics of Wilson loops with corners in some different manner. 
To make the Wilson loops we are interested in well defined in the continuum 
limit, one would need to either use smeared loops, or an alternative method, 
to eliminate the ultraviolet divergences inherent in the field-theoretical 
definition of the observables under consideration. Effective string theory is
supposed to apply to continuum observables and to be free of ultraviolet cutoff effects. 
After applying constraints resulting from spacetime invariance in four dimensions, 
all ambiguities still present are supposedly parameterizable by terms local on 
the worldsheet, on the boundary and at the corners. These extra terms make 
contributions to the Wilson loops that are suppressed
by non-negative powers of the inverse area measured in units of the string tension. 
An effective string theory different from the one employed here might use a different
set of fields, or allow different kinds of additions representing corners, or both.

So far, it seems possible to try to roughly 
match rectangular loops across the large-$N$
phase transition. We need to perform more checks on the short scales side
of the transition to see if one can ultimately turn this into a credible
estimate of the ratio $\sigma/\Lambda_{\overline{MS}}^2$. 

Our results might be taken as an indication to consider other observables that admit an
effective string representation. Much work has been done on the 
two-point correlation function of Polyakov loops. Here, there are 
no corner divergences to worry about. So far one has not established
an analogue of the large-$N$ phase transition in Wilson loops in this
case, but we believe this to be possible. So, maybe focusing on Polyakov loop
correlations would provide a way to temporarily circumvent the issue of
dependence on loop shape in the presence of kinks. 
A more esoteric option is to use surface operators where the regularization of
the operator presents less difficulty \cite{surface-ops}. 

In any case, the issue of shape dependence is seen 
not to be a numerical impediment 
to obtaining a reasonably accurate estimate 
of $\sigma/\Lambda^2$ by analytical means. 
Only refinements at a later stage might 
have to take this issue into account. Nonetheless, 
at a deeper level a full understanding of 
the interplay between field-theoretical properties
of kinks and a good effective string description 
of large loops promises to be illuminating. 

\section{Acknowledgments}

HN would like to thank Ofer Aharony for a very useful 
e-mail exchange and the authors of
~\cite{tony} for a friendly heads-up regarding the 
posting of their preprint. HN thanks the GGI 
Institute for hosting the 2010 workshop on ``Large-$N$ gauge theories''. 
Several discussions
with participants in this workshop were very helpful. 
Our research is supported in part by the 
DOE under grant number DE-FG02-01ER41165. 
 
\clearpage

\section{Tables}

\vspace{1cm}

\begin{table}[htb]
 \begin{tabular}{ | c | c | }
\hline
\multicolumn{2}{|c|}{Method 1)} \\ \hline\hline 
$b$ & $\sigma$ \\ \hline 
0.359 & 0.0492(12) \\ 
0.360 & 0.04597(94) \\ 
0.361 & 0.04170(86) \\ 
0.362 & 0.03946(71) \\ 
0.363 & 0.03767(61) \\ 
0.364 & 0.03544(58) \\ 
0.365 & 0.03266(55) \\ 
0.366 & 0.03105(47) \\ 
0.367 & 0.02878(45) \\ 
0.368 & 0.02701(38) \\ 
0.369 & 0.02508(36) \\  
\hline   
\end{tabular}
\hfill
\begin{tabular}{ | c | c | }
\hline
\multicolumn{2}{|c|}{Method 2a)} \\ \hline\hline 
$b$ & $\sigma$ \\ \hline 
0.359 & 0.0500(22) \\ 
0.360 & 0.0477(20) \\ 
0.361 & 0.0425(15) \\ 
0.362 & 0.0405(14) \\ 
0.363 & 0.0375(12) \\ 
0.364 & 0.0360(10) \\ 
0.365 & 0.0334(10) \\ 
0.366 & 0.03183(87) \\ 
0.367 & 0.02931(89) \\ 
0.368 & 0.02696(76) \\ 
0.369 & 0.02614(69) \\ 
\hline   
\end{tabular}
\hfill
\begin{tabular}{ | c | c | }
\hline
\multicolumn{2}{|c|}{Method 2b)} \\ \hline\hline 
$b$ & $\sigma$ \\ \hline 
\hline   
0.359 & 0.0514(22) \\ 
0.360 & 0.0492(21) \\ 
0.361 & 0.0449(16) \\ 
0.362 & 0.0398(15) \\ 
0.363 & 0.0388(13) \\ 
0.364 & 0.0382(10) \\ 
0.365 & 0.0326(11) \\ 
0.366 & 0.03195(98) \\ 
0.367 & 0.02954(77) \\ 
0.368 & 0.02854(76) \\ 
0.369 & 0.02657(72) \\ 
\hline 
\end{tabular}
\caption{Results for the infinite-$N$ string tension $\sigma$ at $S=0.4$.}
\label{tab:sigmas}
\end{table}


\begin{table}[htb]
 \begin{tabular}{ | c | c | }
\hline
\multicolumn{2}{|c|}{Method 1)} \\ \hline\hline 
$b$ & $\sigma$ \\ \hline 
0.359 & 0.0487(22) \\ 
0.360 & 0.0456(17) \\ 
0.361 & 0.0430(16) \\ 
0.362 & 0.0400(13) \\ 
0.363 & 0.0373(11) \\ 
0.364 & 0.03577(99) \\ 
0.365 & 0.03252(84) \\ 
0.366 & 0.03106(73) \\ 
0.367 & 0.02842(77) \\ 
0.368 & 0.02713(62) \\ 
0.369 & 0.02462(57) \\ 
\hline   
\end{tabular}
\hfill
\begin{tabular}{ | c | c | }
\hline
\multicolumn{2}{|c|}{Method 2a)} \\ \hline\hline 
 $b$ & $\sigma$ \\ \hline 
0.359 & 0.0480(39) \\ 
0.360 & 0.0486(33) \\ 
0.361 & 0.0434(28) \\ 
0.362 & 0.0434(24) \\ 
0.363 & 0.0363(21) \\ 
0.364 & 0.0347(17) \\ 
0.365 & 0.0331(17) \\ 
0.366 & 0.0314(14) \\ 
0.367 & 0.0288(15) \\ 
0.368 & 0.0271(13) \\ 
0.369 & 0.0255(11) \\ 
\hline   
\end{tabular}
\hfill
\begin{tabular}{ | c | c | }
\hline
\multicolumn{2}{|c|}{Method 2b)} \\ \hline\hline 
 $b$ & $\sigma$ \\ \hline 
\hline   
0.359 & 0.0543(40) \\ 
0.360 & 0.0504(37) \\ 
0.361 & 0.0466(30) \\ 
0.362 & 0.0393(25) \\ 
0.363 & 0.0377(22) \\ 
0.364 & 0.0412(18) \\ 
0.365 & 0.0318(19) \\ 
0.366 & 0.0317(16) \\ 
0.367 & 0.0295(14) \\ 
0.368 & 0.0289(13) \\ 
0.369 & 0.0261(12) \\ 
\hline 
\end{tabular}
\caption{Results for the infinite-$N$ string tension $\sigma$ at $S=0.28$.}
\label{tab:sigmasS14}
\end{table}

\TABLE[htb]{
 \begin{tabular}{ | c || c | c || c | c | }
\hline
method \& range & $d_0$ & $\chi^2/N_{\text{dof}}$ & $f_0$ & $\chi^2/N_{\text{dof}}$ \\
\hline 
1) \&  A  & 1.50(6)  & 0.51       & 1.62(3)  & 0.63 \\
2a) \& A  & 1.55(11) & 0.24       & 1.66(6)  & 0.25 \\
2b) \& A  & 1.55(11) & 0.80       & 1.68(5)  & 0.79 \\
1) \&  B  & 1.55(14) & 0.42       & 1.65(7)  & 0.46 \\
2a) \& B  & 1.64(27) & 0.14       & 1.71(16) & 0.14 \\
2b) \& B  & 1.54(26) & 1.56       & 1.68(13) & 1.59 \\
\hline
\end{tabular}
\caption{Extrapolation to the continuum using $\sigma$ from Table~\ref{tab:sigmas}.}
\label{tab:ContRes}
}

\TABLE[htb]{
 \begin{tabular}{ | r || c | c || c | c | }
\hline
fit & $d_0$ & $\chi^2/N_{\text{dof}}$ & $f_0$ & $\chi^2/N_{\text{dof}}$ \\
\hline 
1) \&  A  & 1.89(8) &0.51 &    2.10(3) & 0.69 \\
2a) \& A  & 1.94(15) &0.24 &   2.15(7) & 0.26\\
2b) \& A  & 1.95(16) & 0.80 &  2.18(7) & 0.79  \\
1) \&  B  & 1.95(19) &0.42 &   2.13(9) & 0.48 \\
2a) \& B  & 2.07(37) &0.14 &   2.20(19) &0.15 \\
2b) \& B  & 1.93(36) &1.56 &   2.17(15) & 1.61 \\
\hline
\end{tabular}
\caption{Continuum extrapolations using $\xi_c(b)$ with setting $\bar\beta_2=0$ in \eqref{eq:Lc}.}
\label{tab:ContResBeta2NULL}
}

\TABLE[htb]{
\footnotesize{
 \begin{tabular}{ | c || c | c | c | c || c | c | c | }
\hline
$b$ & $\sigma_7$ & $\sigma_{11}$ & $\sigma_{19}$ & $\sigma_{29}$ &
$\sigma_\infty$ & $h$ & $\chi^2/N_{\text{dof}}$ \\
\hline
\hline
0.359 & 0.03413(38) & 0.04337(65) & 0.04726(79) & 0.0480(11) & 0.04927(60) &
-0.741(37) & 0.14\\
0.360 & 0.03219(33) & 0.03924(57) & 0.04418(71) & 0.04476(87) & 0.04533(53) &
-0.647(33) & 1.3\\
0.361 & 0.03086(32) & 0.03804(54) & 0.04094(60) & 0.04100(74) & 0.04231(46) &
-0.559(29) & 0.64 \\
0.362 & 0.02904(28) & 0.03546(43) & 0.03839(52) & 0.03888(66) & 0.03974(39) &
-0.524(26) & 0.09 \\
0.363 & 0.02717(27) & 0.03318(40) & 0.03673(44) & 0.03678(54) & 0.03772(34) &
-0.518(23) & 0.91 \\
0.364 & 0.02599(23) & 0.03168(37) & 0.03449(37) & 0.03438(55) & 0.03551(31) &
-0.466(20) & 0.81 \\
0.365 & 0.02420(24) & 0.02916(35) & 0.03124(40) & 0.03240(47) & 0.03259(30) &
-0.412(20) & 0.35 \\
0.366 & 0.02314(21) & 0.02781(30) & 0.03033(38) & 0.03032(41) & 0.03108(27) &
-0.389(18) & 0.68 \\
0.367 & 0.02172(18) & 0.02610(28) & 0.02834(28) & 0.02810(40) & 0.02906(23) &
-0.359(16) & 1.3 \\
0.368 & 0.02032(17) & 0.02468(27) & 0.02677(28) & 0.02647(32) & 0.02740(21) & 
-0.345(14) & 2.2 \\
0.369 & 0.01917(16) & 0.02361(25) & 0.02510(25) & 0.02473(31) & 0.02582(20) &
-0.322(13) & 4.6 \\
\hline
\end{tabular}
}
\caption{String tension at finite $N=7$, $11$, $19$, $29$, and results of the
  corresponding extrapolations to infinite $N$ (obtained by fitting
  $\sigma_\infty$ and $h$ in Eq.~\eqref{eq:sigmaFinNCorr}).}
\label{tab:sigmaN}
}

\begin{table}
\begin{center}
 \begin{tabular}{ | c | c | }
\hline
\multicolumn{2}{|c|}{Method 1) for rect.} \\ \hline\hline 
$b$ & $\sigma$ \\ \hline 
0.359 & 0.04842(66) \\ 
0.360 & 0.04510(56) \\ 
0.361 & 0.04153(53) \\ 
0.362 & 0.03943(43) \\ 
0.363 & 0.03744(45) \\ 
0.364 & 0.03486(39) \\ 
0.365 & 0.03239(37) \\ 
0.366 & 0.03068(29) \\ 
0.367 & 0.02858(29) \\ 
0.368 & 0.02688(25) \\ 
0.369 & 0.02489(26) \\ 
\hline   
\end{tabular}
\caption{Results for the infinite-$N$ string tension $\sigma$ from rectangular
  $L\times L\pm1$ loops at $S=0.4$.}
\label{tab:sigmasRCT}
\end{center}
\end{table}

\TABLE[htb]{
\footnotesize{
 \begin{tabular}{ | c | c || l | l | l | l | l || c | }
\hline 
$b$ & $S$ & \multicolumn{1}{|c|}{$\sigma$} & \multicolumn{1}{c|}{$c_3$} & \multicolumn{1}{c|}{$c_2$} & \multicolumn{1}{c|}{$c_1$} & \multicolumn{1}{c||}{$c_4$} & $\chi^2/N_{\text{dof}}$ \\
\hline
\hline
 0.365 & 0.52  & 0.02462(29) & -0.2415(72) & 0.3739(55) & -0.4423(71) & -0.2700(40) & 0.25 \\ 
 0.365 & 0.44  & 0.02418(33) & -0.2558(84) & 0.4332(64) & -0.4895(81) & -0.2819(46) & 0.31 \\ 
 0.365 & 0.36  & 0.02379(38) & -0.267(10) & 0.5066(75) & -0.5478(93) & -0.2918(54) & 0.42 \\ 
 0.365 & 0.28  & 0.02348(48) & -0.273(13) & 0.6028(95) & -0.622(11) & -0.2991(67) & 0.45 \\ 
 0.365 & 0.20  & 0.02348(73) & -0.269(19) & 0.734(14) & -0.714(15) & -0.3011(100) & 0.77 \\ 
 \hline
 0.366 & 0.52  & 0.02364(29) & -0.2349(70) & 0.3696(55) & -0.4414(59) & -0.2657(40) & 0.20 \\ 
 0.366 & 0.44  & 0.02333(33) & -0.2444(83) & 0.4249(65) & -0.4906(67) & -0.2752(47) & 0.15 \\ 
 0.366 & 0.36  & 0.02315(39) & -0.249(10) & 0.4928(78) & -0.5510(78) & -0.2816(57) & 0.23 \\ 
 0.366 & 0.28  & 0.02307(49) & -0.249(13) & 0.5832(97) & -0.6265(98) & -0.2853(70) & 0.48 \\ 
 0.366 & 0.20  & 0.02267(69) & -0.252(17) & 0.721(13) & -0.721(14) & -0.2920(96) & 0.75 \\ 
 \hline
 0.367 & 0.52  & 0.02189(25) & -0.2377(64) & 0.3775(49) & -0.4520(60) & -0.2707(35) & 0.27 \\ 
 0.367 & 0.44  & 0.02155(29) & -0.2488(76) & 0.4331(57) & -0.4988(68) & -0.2808(41) & 0.15 \\ 
 0.367 & 0.36  & 0.02133(34) & -0.2565(92) & 0.5018(68) & -0.5549(80) & -0.2882(49) & 0.14 \\ 
 0.367 & 0.28  & 0.02125(43) & -0.260(12) & 0.5926(89) & -0.6226(97) & -0.2928(63) & 0.20 \\ 
 0.367 & 0.20  & 0.02122(64) & -0.262(17) & 0.724(13) & -0.706(13) & -0.2961(91) & 0.24 \\ 
 \hline
 0.368 & 0.52  & 0.02073(24) & -0.2392(63) & 0.3801(47) & -0.4529(55) & -0.2720(34) & 0.34 \\ 
 0.368 & 0.44  & 0.02030(27) & -0.2516(75) & 0.4367(55) & -0.5004(64) & -0.2832(40) & 0.19 \\ 
 0.368 & 0.36  & 0.01990(33) & -0.2616(90) & 0.5075(66) & -0.5582(76) & -0.2928(48) & 0.16 \\ 
 0.368 & 0.28  & 0.01951(41) & -0.271(11) & 0.6028(84) & -0.6283(91) & -0.3017(60) & 0.28 \\ 
 0.368 & 0.20  & 0.01905(58) & -0.280(16) & 0.740(12) & -0.713(12) & -0.3107(83) & 0.67 \\ 
 \hline
 0.369 & 0.52  & 0.01996(21) & -0.2195(57) & 0.3696(42) & -0.4667(53) & -0.2637(30) & 0.17 \\ 
 0.369 & 0.44  & 0.01968(25) & -0.2275(70) & 0.4224(50) & -0.5152(60) & -0.2726(36) & 0.12 \\ 
 0.369 & 0.36  & 0.01945(30) & -0.2322(86) & 0.4886(62) & -0.5746(70) & -0.2794(44) & 0.19 \\ 
 0.369 & 0.28  & 0.01921(38) & -0.236(11) & 0.5793(78) & -0.6474(82) & -0.2855(56) & 0.21 \\ 
 0.369 & 0.20  & 0.01872(52) & -0.244(14) & 0.716(11) & -0.734(11) & -0.2946(75) & 0.29 \\ 
 \hline
\hline
\end{tabular}
}
\caption{Fit results for $N=7$ on $V=24^4$. Loop sizes used for the fits are $5\times5$ to $11\times11$,
$5\times6$ to $11\times 12$, $4\times 8$ to $8\times 16$ (which results in $N_{\text{dof}}=14$)}
\label{tab:fullfitN7}
}

\TABLE[htb]{
\footnotesize{
 \begin{tabular}{ | c | c || l | l | l | l | l || c | }
\hline
$b$ & $S$ & \multicolumn{1}{|c|}{$\sigma$} & \multicolumn{1}{c|}{$c_3$} & \multicolumn{1}{c|}{$c_2$} & \multicolumn{1}{c|}{$c_1$} & \multicolumn{1}{c||}{$c_4$} & $\chi^2/N_{\text{dof}}$ \\
\hline
\hline
 0.359 & 0.52  & 0.04347(77) & -0.233(15) & 0.334(13) & -0.421(14) & -0.2573(95) & 0.41 \\ 
 0.359 & 0.36  & 0.04167(99) & -0.284(20) & 0.505(18) & -0.540(18) & -0.294(13) & 0.31 \\ 
 \hline
 0.360 & 0.52  & 0.03912(74) & -0.262(16) & 0.371(14) & -0.444(13) & -0.282(10) & 0.26 \\ 
 0.360 & 0.36  & 0.03836(97) & -0.284(22) & 0.518(18) & -0.573(19) & -0.304(13) & 0.28 \\ 
 \hline
 0.361 & 0.52  & 0.03792(75) & -0.241(16) & 0.355(14) & -0.442(12) & -0.2686(100) & 0.08 \\ 
 0.361 & 0.36  & 0.03694(98) & -0.264(22) & 0.503(18) & -0.573(17) & -0.293(13) & 0.23 \\ 
 \hline
0.362 & 0.52  & 0.03531(58) & -0.260(14) & 0.369(11) & -0.424(12) & -0.2756(79) & 0.25 \\ 
 0.362 & 0.36  & 0.03456(78) & -0.281(19) & 0.512(15) & -0.547(16) & -0.297(11) & 0.22 \\ 
 0.362 & 0.20  & 0.0335(14) & -0.308(34) & 0.777(27) & -0.716(26) & -0.319(19) & 0.16 \\ 
 \hline
 0.363 & 0.52  & 0.03267(55) & -0.270(13) & 0.384(10) & -0.435(11) & -0.2849(75) & 0.41 \\ 
 0.363 & 0.36  & 0.03157(73) & -0.302(18) & 0.533(14) & -0.552(14) & -0.312(10) & 0.63 \\ 
 0.363 & 0.20  & 0.0319(13) & -0.289(31) & 0.766(25) & -0.734(24) & -0.313(18) & 0.69 \\ 
 \hline
 0.364 & 0.52  & 0.03185(49) & -0.241(13) & 0.3652(97) & -0.4440(99) & -0.2697(70) & 0.12 \\ 
 0.364 & 0.36  & 0.03102(66) & -0.263(17) & 0.505(13) & -0.564(13) & -0.2912(94) & 0.17 \\ 
 0.364 & 0.20  & 0.0301(13) & -0.273(33) & 0.758(26) & -0.750(22) & -0.309(18) & 0.33 \\ 
 \hline
 0.365 & 0.52  & 0.02929(44) & -0.257(11) & 0.3815(85) & -0.4432(96) & -0.2795(62) & 0.25 \\ 
 0.365 & 0.36  & 0.02837(59) & -0.278(15) & 0.521(12) & -0.565(13) & -0.3021(83) & 0.15 \\ 
 0.365 & 0.20  & 0.0280(11) & -0.274(29) & 0.759(22) & -0.752(20) & -0.312(16) & 0.31 \\ 
 \hline
 0.366 & 0.52  & 0.02816(46) & -0.251(11) & 0.3756(87) & -0.4357(90) & -0.2731(63) & 0.27 \\ 
 0.366 & 0.36  & 0.02745(58) & -0.276(15) & 0.512(11) & -0.542(12) & -0.2940(81) & 0.57 \\ 
 0.366 & 0.20  & 0.02689(98) & -0.287(24) & 0.755(19) & -0.710(21) & -0.307(13) & 0.89 \\ 
 \hline
 0.367 & 0.52  & 0.02638(39) & -0.2388(96) & 0.3763(76) & -0.4611(85) & -0.2733(55) & 0.18 \\ 
 0.367 & 0.36  & 0.02557(55) & -0.258(14) & 0.510(11) & -0.578(11) & -0.2940(77) & 0.12 \\ 
 0.367 & 0.20  & 0.02519(92) & -0.254(23) & 0.743(18) & -0.757(19) & -0.302(13) & 0.31 \\ 
 \hline
 0.368 & 0.52  & 0.02504(37) & -0.2378(94) & 0.3773(71) & -0.4614(84) & -0.2730(52) & 0.12 \\ 
 0.368 & 0.36  & 0.02421(50) & -0.261(13) & 0.5109(98) & -0.572(11) & -0.2943(71) & 0.04 \\ 
 0.368 & 0.20  & 0.02361(84) & -0.266(23) & 0.747(17) & -0.745(19) & -0.307(12) & 0.38 \\ 
 \hline
 0.369 & 0.52  & 0.02391(34) & -0.2492(92) & 0.3796(68) & -0.4335(79) & -0.2720(49) & 0.29 \\ 
 0.369 & 0.36  & 0.02321(45) & -0.270(12) & 0.5094(90) & -0.540(11) & -0.2910(65) & 0.38 \\ 
 0.369 & 0.20  & 0.02251(81) & -0.269(21) & 0.742(16) & -0.723(17) & -0.302(11) & 0.45 \\ 
 \hline
\hline
\end{tabular}
}
\caption{Fit results for $N=11$ on $V=18^4$. Loop sizes used for the fits are $5\times5$ to $11\times11$,
$5\times6$ to $11\times 12$, $4\times 8$ to $8\times 16$.}
\label{tab:fullfitN11}
}

\end{document}